\documentclass[onecolumn,prd,a4paper]{revtex4-1}

\usepackage{amsmath}
\usepackage{graphicx}
\usepackage{psfrag}
\usepackage{color}

\newcommand{\bq}{\begin{equation}}
\newcommand{\eq}{\end{equation}}
\newcommand{\ba}{\begin{eqnarray}}
\newcommand{\ea}{\end{eqnarray}}

\begin{document}

\title{Quantum widening of CDT universe}
\author{L. Bogacz}
\affiliation{Department of Information Technologies, Faculty of Physics, Astronomy and Applied Informatics, Jagellonian University, 
Reymonta 4, 30-059 Krakow, Poland.}
\author{Z. Burda}
\affiliation{Marian Smoluchowski Institute of Physics, Jagellonian University, Reymonta 4, 30-059 Krak\'ow, Poland}
\author{B. Waclaw}
\affiliation{School of Physics and Astronomy, University of Edinburgh, Mayfield Road, Edinburgh EH9 3JZ, United Kingdom}

\begin{abstract}
The physical phase of Causal Dynamical Triangulations (CDT) is known to be described by an effective, one-dimensional action in which three-volumes of the underlying foliation of the full CDT play a role of the sole degrees of freedom. Here we map this effective description onto a statistical-physics model of particles distributed on 1d lattice, with site occupation numbers corresponding to the three-volumes. We identify the emergence of the quantum de-Sitter universe observed in CDT with the condensation transition known from similar statistical models.
Our model correctly reproduces the shape of the quantum universe and allows us to analytically determine quantum corrections to the size of the universe. We also investigate the phase structure of the model and show that it reproduces all three phases observed in computer simulations of CDT. In addition, we predict that two other phases may exists, depending on the exact form of the discretised effective action and boundary conditions. We calculate various quantities such as the distribution of three-volumes in our model and discuss how they can be compared with CDT.
\end{abstract}

\maketitle

\section{Introduction}
Causal dynamical triangulations (CDT) \cite{cdt1,cdt2,cdt3} is an attempt to construct a non-perturbative theory of quantum gravity. Rather than postulating the existence of new degrees of freedom or new physical principles at the Planck scale, CDT uses a standard quantum field theory method --- path integrals --- to sum over space-time geometries weighted by the Einstein-Hilbert action. The path integrals are regularised by discretisation of space-time geometry into piece-wise flat manifolds with temporal foliation.
Usually, space-time is divided into discrete spatial slices, each having the topology of the three-sphere, which ensures global, proper-time foliation consistent with the Lorentzian signature of the metric. Each spatial slice is represented as a triangulation of the three-sphere, made of equilateral tetrahedra. The tetrahedra from neighbouring spatial slices are then glued together, thus forming a complicated 4d manifold, with periodic boundary conditions in time direction. This lattice regularisation provides a suitable ultraviolet cut-off and simultaneously reproduces classical general relativity in the infrared limit.

Although analytic calculations do not seem to be feasible in the full 3+1 dimensional CDT, the model can be studied by means of computer simulations. After the Wick rotation to the Euclidean signature, the sum over geometries can be performed by standard Monte Carlo methods developed earlier for Euclidean quantum gravity \cite{eqt,am,bs,bbkp}. In recent years, it has been shown that this computational approach has a potential to bring many interesting results. In particular, the existence of three phases has been observed \cite{phases1}. These phases have different profiles of the three-volume $N_3(t)$ as a function of time (slice index) $t$. Depending on the values of parameters in the Einstein-Hilbert action, the system is either in phase ``A'', in which $N_3(t)$ fluctuates randomly from slice to slice, phase ``B'' in which $N_3(t)$ is localised in a single spatial slice, or in phase ``C'' in which a macroscopic ``quantum universe'' is formed  \cite{quantum1,quantum2,quantum3,quantum4}. In this last phase, the average value of $N_3(t)$ at each time slice $t$ is well described by the following formula:
\bq
\langle N_3(t)\rangle = \left\{ \begin{array}{ll} \frac{N_4^{3/4}}{2s} \cos^3\left(\frac{t}{s N_4^{1/4}} \right) & \mbox{for}\; |t|<\frac{\pi s N_4^{1/4}}{2}, \\
0 & \mbox{for}\; |t|\geq \frac{\pi s N_4^{1/4}}{2},
\end{array} \right. \label{cos3}
\eq
where $N_4=\sum_t N_3(t)$ is the total (fixed) four-volume of the universe;
$s$ is obtained by fitting to the results of simulations; the centre of mass is assumed to be at $t=0$. The last formula means that the universe produces a ``droplet'' of $\cos^3(x)$ shape, and that this droplet extends as $\pi s N_4^{1/4}$ in time direction. This shape
is equivalent to the classical de-Sitter solution.
By making a connection with the mini-superspace model \cite{hh} it has been concluded in Refs. \cite{effact1,phases1,quantum2,quantum3,quantum4} that, when only the three-volume is concerned, the full CDT model effectively reduces to a 1d model with three-volumes $\{N_3(t)\}$ as the sole degrees of freedom, and with the following discrete action:
\bq
	S = c_1 \sum_t \frac{(N_3(t+1)-N_3(t))^2}{N_3(t)} + c_2 \sum_t N_3^{1/3}(t), \label{seff}
\eq
Here $c_1,c_2$ are new coupling constants  related to those in the full Einstein-Hilbert action. An important fact is that although the action (\ref{seff}) completely neglects the internal structure of each spatial slice $t$, it gives an excellent agreement with simulations of the full model.

In this paper we introduce a statistical-physics model which reproduces the de-Sitter phase of the CDT. Our model consist of a certain number of particles which occupy sites of a 1d lattice, and microstates (configurations of particles) are weighted with the factor $e^{-S}$. We identify the emergence of the de-Sitter universe with a condensation-like transition known from similar statistical models \cite{evans1,bw1}. We show (both analytically and via computer simulations) that a symmetrised version of the action (\ref{seff}) reproduces the shape of the macroscopic universe observed in CDT. We calculate the width (temporal extension) of this universe and show that quantum corrections make it wider as compared to the classical solution. 

Moreover, we show that the effective action (\ref{seff}) describes not only phase C of CDT but also phases A and B, in the space of the coupling constants $c_1,c_2$. In addition, we suggest that two further phases may exist: ``antiferromagnetic'' phase D in which thin spatial slices of extended three-volume are separated by slices of minimal size, and ``correlated fluid'' phase E which emerges from phase C for large four-volume $N_4$ as a result of merging boundaries of the $\cos^3(x)$-shaped universe. In all these phases we calculate quantities such as the probability distribution of the three-volume or the correlation function for different three-volumes. Lastly, we suggest that by determining analogous quantities in CDT it should be possible to test whether the effective action (\ref{seff}) is valid in all phases.

\section{Model}
In our model, we consider a one-dimensional closed ring of $N$ sites, each of them carrying a positive number of particles $m_1\geq 1,\dots,m_N\geq 1$. The total number of particles is equal to $M$. We denote the density of particles by $\rho=M/N$. The numbers of sites $N$ and particles $M$ correspond to the numbers of spatial slices and four-volume $N_4$, respectively, while the occupation numbers $\{m_i\}$ correspond to three-volumes $\{N_3(t)\}$ of spatial slices in CDT. 

We assume that the probability of a microstate $P(m_1,\dots,m_N)$ factorizes into the product of two-point kernels for pairs of neighbouring sites,
\bq
P(m_1,\dots,m_N) = g(m_1,m_2) g(m_2,m_3)...g(m_{N-1},m_N)g(m_N,m_1), \label{Pss}
\eq
where 
\bq
g(m,n) = \exp\left(-c_1 \frac{2(m-n)^2}{m+n} -c_2 \frac{m^{1/3}+n^{1/3}}{2}\right) \ .\label{gmn}
\eq
The kernel $g(m,n)$ plays the role of a reduced transfer matrix between neighbouring slices of CDT. The above choice guarantees that the partition function
\ba
Z(N,M) &=& \sum_{m_1=1}^M ... \sum_{m_N=1}^M g(m_1,m_2) g(m_2,m_3)...g(m_{N-1},m_N)g(m_N,m_1) \delta\left(\sum_i m_i-M\right)
\nonumber \\
&= & \sum_{m_1=1}^M ... \sum_{m_N=1}^M \exp\left[ - \sum_i \left(c_1\frac{(m_{i+1}-m_i)^2}{(m_i+m_{i+1})/2} + c_2 m_i^{1/3}\right)\right] \delta\left(\sum_i m_i-M\right) \nonumber \\ &= & \sum_{m_1=1}^M ... \sum_{m_N=1}^M \exp\big[ -S\left[\{m_i\}\right]\big] \delta\left(\sum_i m_i-M\right) \label{zdef}
\ea
corresponds to that of CDT with the effective action (\ref{seff}) in the limit of large systems. Our choice (\ref{gmn}) is however symmetric in $n,m$ as opposed to (\ref{seff}). We shall see later that this symmetry is necessary to reproduce full-CDT simulation results.

Equation (\ref{Pss}) has the same form as the steady-state probability of a recently introduced non-equilibrium statistical physics model of particles hopping between sites of a 1d lattice \cite{evans1,bw1}. A key feature of this model is the condensation phenomenon in which a finite fraction of particles becomes localised in a small region of the lattice if the density of particles $\rho=M/N$ exceeds some critical value $\rho_c$. In particular, in Ref.~\cite{bw1} the following two-point function $g(m,n)$ has been analysed:
\bq
	g(m,n)=K(|m-n|)\sqrt{f(m)f(n)}, \label{BWgmn}
\eq
with two functions $K(x),f(m)$ playing the role of surface stiffness and on-site potential, respectively. This model has a rich phase diagram which depends on the choice of $K(x)$ and $f(m)$. We will briefly discuss some results of Ref. \cite{bw1} because they are important for the model discussed in this paper. Let us begin with defining the grand-canonical partition function
\bq
Z_N(z) = \sum_M Z(N,M) z^M = \sum_{\{m_i\}} z^{\sum_i m_i} \prod_i g(m_i,m_{i+1}),
\label{zgrand}
\eq
in which the fugacity $z$ is determined from
\bq
\rho =\frac{1}{N}\left<\sum_i m_i\right> = \frac{z}{N}\frac{\partial\ln Z_N(z)}{\partial z} . \label{rhoz}
\eq
We note that the left-hand side of Eq.~(\ref{rhoz}) grows monotonously with $z$. Since phase transitions are related to singularities of 
$Z(z)= \lim_{N\rightarrow\infty} Z_N(z)$ and its derivatives, we are interested in the behaviour of this function as $z$ approaches the radius of convergence $z_c$ of $Z(z)$. If $z_c$ is infinite, there is always some $z>0$ which obeys Eq.~(\ref{rhoz}) for any $\rho$. This means that $Z(z)$ does not have a singularity for $0<z<\infty$ which is the physically relevant range of $z$. Also, both ensembles, the canonical and the grand-canonical one, are equivalent in the thermodynamic limit in this case. The partition function $Z_N(z)$ can be expressed as
\bq
Z_N(z) = \sum_{m_1,\dots,m_N} T_{m_1 m_2} T_{m_2 m_3} \cdots T_{m_N m_1} = \mbox{Tr}\, T(z)^N, \label{zcrit}
\eq
where $T(z)$ is a square $M\times M$ matrix defined as
\bq
T_{mn}(z)=z^{(m+n)/2}g(m,n). \label{eq:tmn}
\eq
If we now define $\phi_m(z)$ to be the normalised eigenvector of $T_{mn}(z)$ to the largest eigenvalue $\lambda_{\rm max}(z)$,
\bq
	\sum_n T_{mn}(z) \phi_n(z) = \lambda_{\rm max}(z) \phi_m(z) ,\label{eq:eigen}
\eq
we obtain for large $N$ that $Z_N(z) \cong \lambda_{\rm max}(z)^N$. We can also calculate the probability $p(m)$ that a randomly chosen site has $m$ particles:
\bq
	p(m) = \lim_{N\rightarrow \infty} \frac{1}{Z_N(z)} \sum_{m_2,\dots,m_N} T_{m m_2} T_{m_2 m_3} \cdots T_{m_N m} = \phi_m^2(z), \\
\eq
The eigenvector $\phi_m(z)$ decays with $m$, and so does $p(m)$. This is guaranteed by the fact that $\rho$ from Eq.~(\ref{rhoz}) is finite. In this case, the system has a finite number of particles at every site -- we say that the system is in the ``fluid'' phase. One can also show that there are only local correlations between different $m_i$'s in this phase. We shall therefore call this phase a weakly-correlated fluid.

On the other hand, if $Z(z) = \lim_{N\rightarrow \infty} Z_N(z)$ has some finite radius of convergence $z_c<\infty$, the derivative in (\ref{rhoz}) can either grow to infinity for $z\to z_c$, or tend to a finite constant. In the first case, we again have no phase transition, because for any $\rho$ there exists some real $z<z_c$ which obeys Eq.~(\ref{rhoz}). However, if $z_c<\infty$ and $dZ(z)/dz|_{z\to z_c}\to \rm const$, there is a critical density
\bq
	\rho_c = \sum_m m \phi_m(z_c)^2, \label{rhoc}
\eq
above which the grand-canonical ensemble does not exist. This indicates a phase transition from the fluid to the condensed state.

The nature of the condensate depends on $K(x)$ and $f(m)$ which define $g(m,n)$. If $K(x)=\rm const$ and $f(m)$ falls off sufficiently fast, the condensate spontaneously forms on one randomly chosen site, breaking translational invariance. This is precisely the balls-in-boxes (B-in-B) \cite{b-in-b} or zero-range process (ZRP) \cite{evans0501338} condensation. We shall note here that the B-in-B model has already been successfully applied to the transition between crumpled and elongated phase in Euclidean quantum gravity models \cite{b-in-b-euclid,b-in-b-eff}.

If $K(x)$ decays with $x$, the condensate extends to more than one site. The width $W$ of the condensate grows as some power $\alpha$ of its volume, $W\sim M^\alpha$. The condensate can be either bell-shaped, or rectangular, depending on exact forms of $K(x)$ and $f(m)$. We see that this closely resembles the features of the macroscopic universe from phase C in CDT. This type of phase, which we shall call a ``droplet'' phase, will be discussed extensively in Section \ref{sec:droplet}. However, if the extension $W$ of the condensate becomes comparable to the linear extension of the system $N$, both ends of the condensate merge and the particles spread uniformly in the system. This phase differs from the weakly-correlated fluid phase which exists for $\rho<\rho_c$ in that the occupation numbers $\{m_i\}$ are correlated. We shall call this phase, rather obviously, a ``correlated fluid'' phase.

It is important to note that the existence of these phases does not depend on the details of $K(x)$ and $f(m)$ (\ref{BWgmn}), which often only affect the shape of the condensate and the value of the critical density. In what follows we shall use the analogy between this model and the effective 1d model of CDT to study the emergence of the bell-shaped quantum universe. We shall also investigate the phase diagram of the model, assuming that the effective action is valid in all phases. A small difference between our model and the model from Refs.~\cite{evans1,bw1} is that the two-point kernel $g(m,n)$ of the 1d effective CDT model (as given in Eq.~(\ref{gmn})) has a slightly different form that Eq.~(\ref{BWgmn}), because $K(x)$ depends not only on the difference between two consecutive occupation numbers $m,n$ but also on their absolute magnitudes $m,n$:
\bq
g(m,n)=K(|m-n|/\sqrt{m+n})\sqrt{f(m)f(n)}
\eq
However, as we shall see, the only new result is the existence of the antiferromagnetic phase which is not observed in the model with kernels of the form (\ref{BWgmn}).

\section{Phase diagram}
We begin with presenting the results of Monte Carlo simulations of our model (see Appendix for details), which reveal its rich phase structure. Anticipating the existence of condensed/fluid phase, and also extended/localised condensates, we define the following quantities which allow us to detect phase transitions: 
\ba
\sigma &=& \left\langle \frac{\sum_i^N m_i^2}{M^2} \right\rangle , \\
\gamma &=& 1 - \left\langle \frac{1}{\min m_i} \right\rangle , \\
\delta &=& 1 - \left\langle \frac{\sum_i^N |m_i - m_{i+1}|}{2M} \right\rangle .
\ea
These quantities assume values between 0 and 1 and play the role of order parameters in the limit $N,M\rightarrow \infty$. The parameter $\sigma$ is the inverse participation ratio for site occupation numbers and it measures the degree of localisation: for a delocalised microstate in which all $m_i$'s are roughly the same, $\sigma\approx 1/N \rightarrow 0$ for $N\rightarrow \infty$. However, if one $m_i$ is much larger than others, $\sigma \rightarrow 1$. The value of the parameter $\gamma$ indicates whether there are any sites with minimal number of particles $m_i=1$ in typical configurations (slices with the smallest possible three-volume in CDT): $\gamma = 0$ if there are such sites, whereas $\gamma>0$ if all sites are occupied by larger numbers of particles. The parameter $\delta$ is related to the surface roughness or stiffness of typical configurations and is close to zero for configurations in which $\{m_i\}$ dramatically change from site to site, and close to one for relatively smooth
configurations.  

\begin{figure}
\centering
\includegraphics[width=11cm]{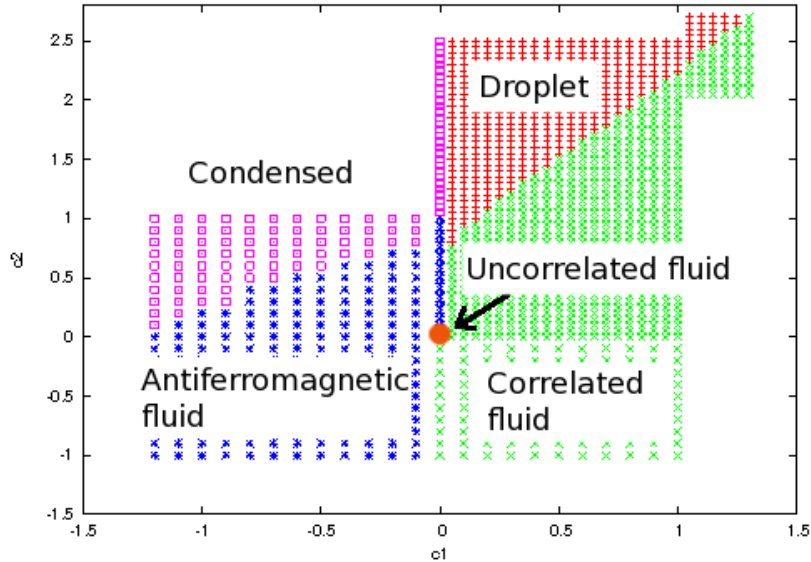} 
\caption{\label{phases}Phase diagram determined from Monte Carlo simulations for $N=80$ and $M=18100$.}
\end{figure}

\begin{figure}
\centering
\psfrag{x}{$c_2$} \psfrag{y}{$\sigma$}
\includegraphics[width=5cm]{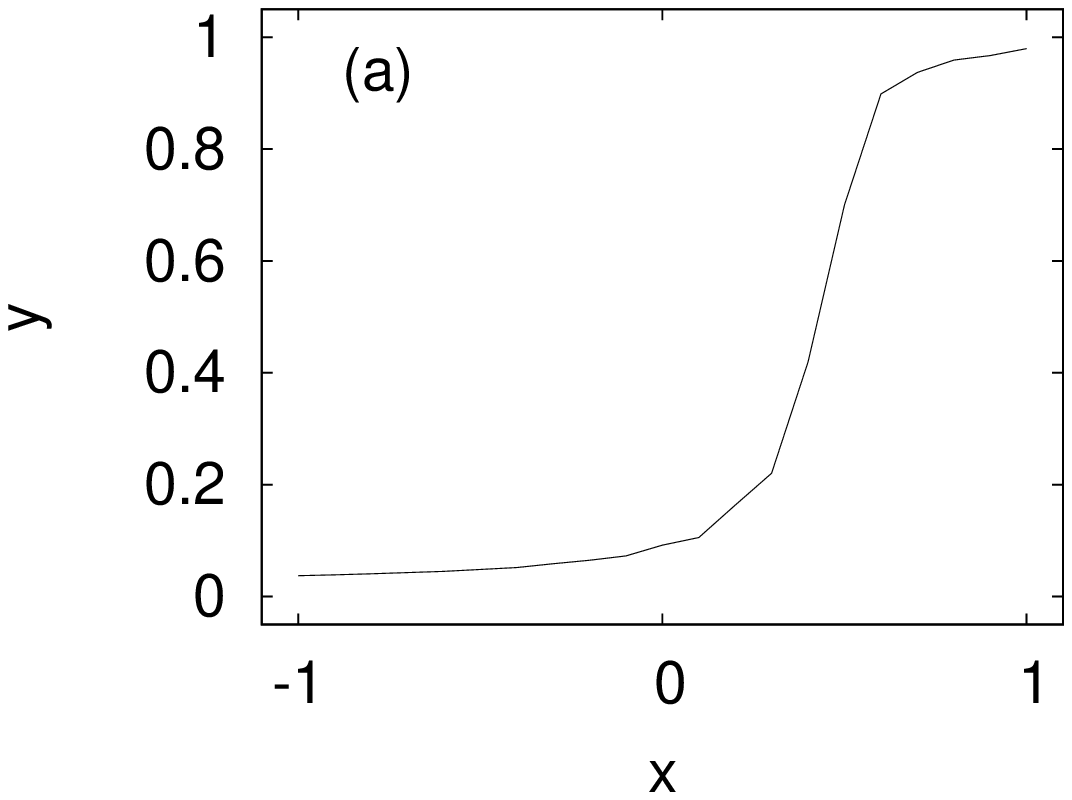}
\psfrag{y}{$\gamma$}
\includegraphics[width=5cm]{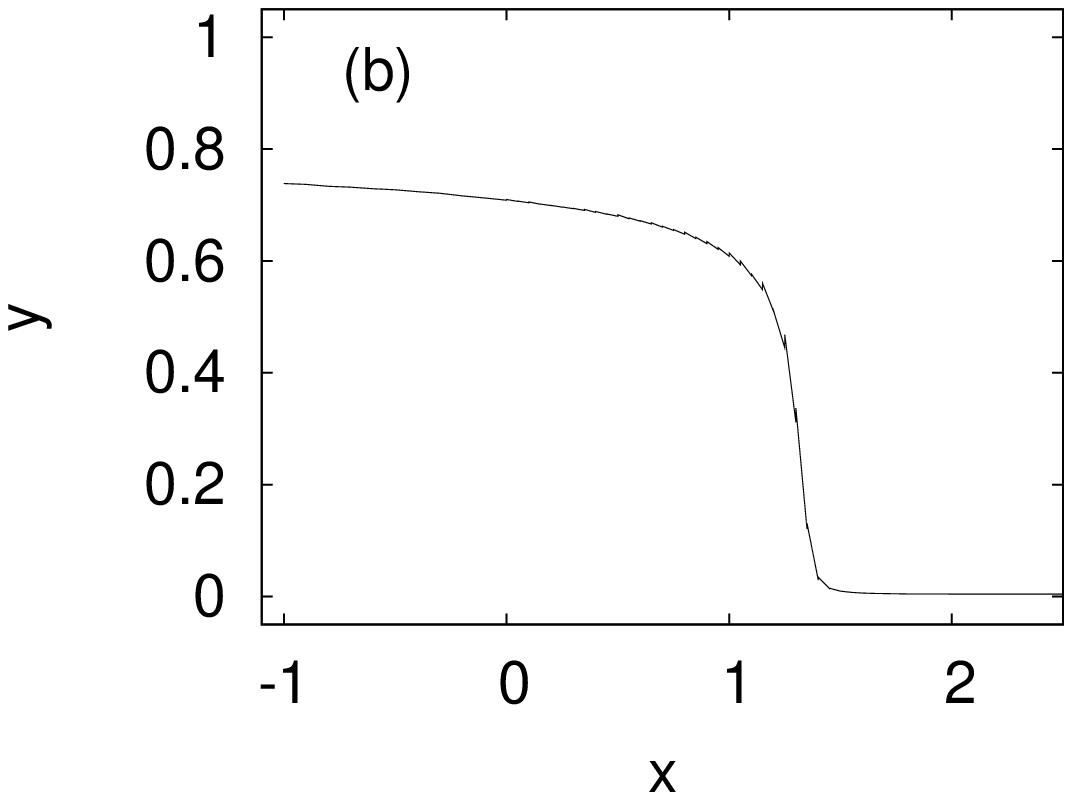}
\psfrag{x}{$c_1$} \psfrag{y}{$\delta$}
\includegraphics[width=5cm]{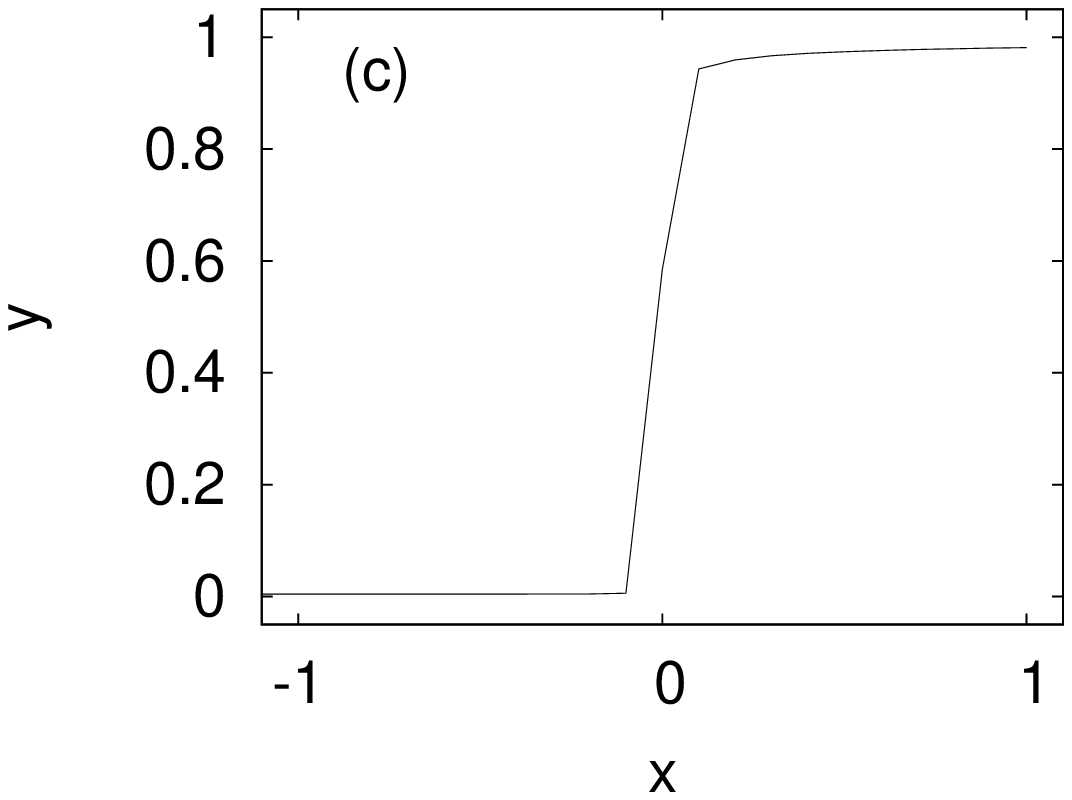}
\caption{\label{phases2}Order parameters: (a) $\sigma(c_2)$ for $c_1=-0.8$, (b) $\gamma(c_2)$ for $c_1=0.5$ and (c) $\delta(c_1)$ for $c_2=-0.5$.}
\end{figure}

\begin{figure}
\psfrag{x}{$i$} \psfrag{y}{$m_i$}
\includegraphics[width=3.5cm]{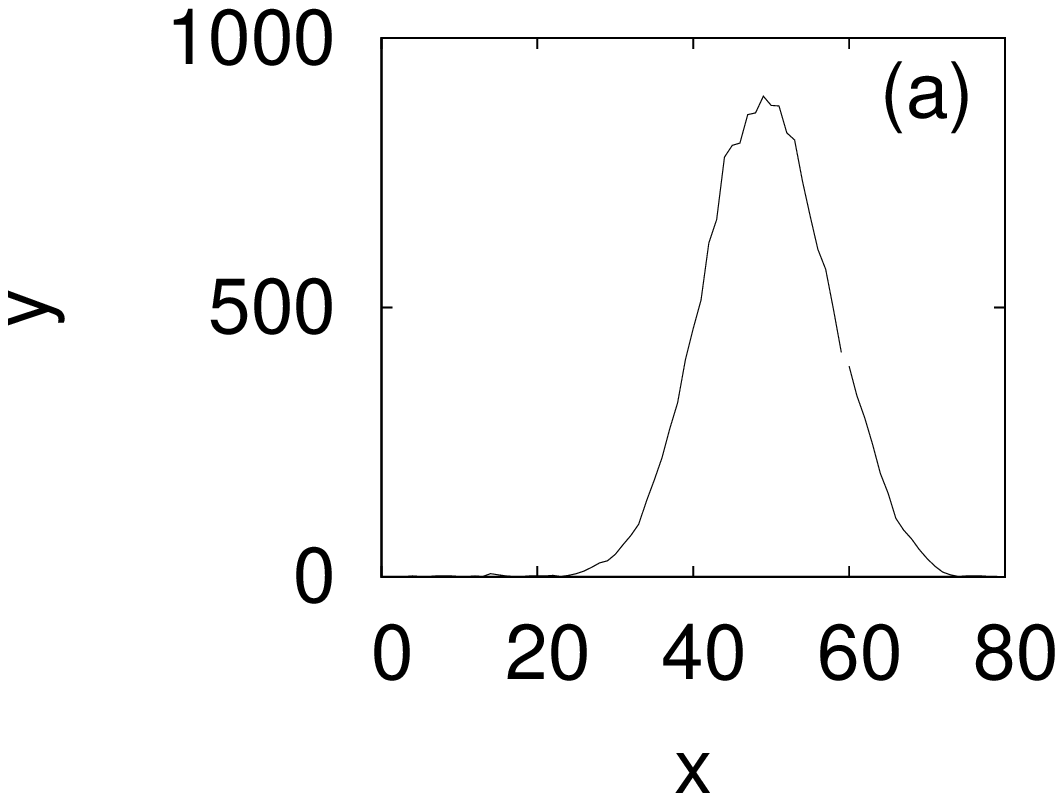}
\includegraphics[width=3.5cm]{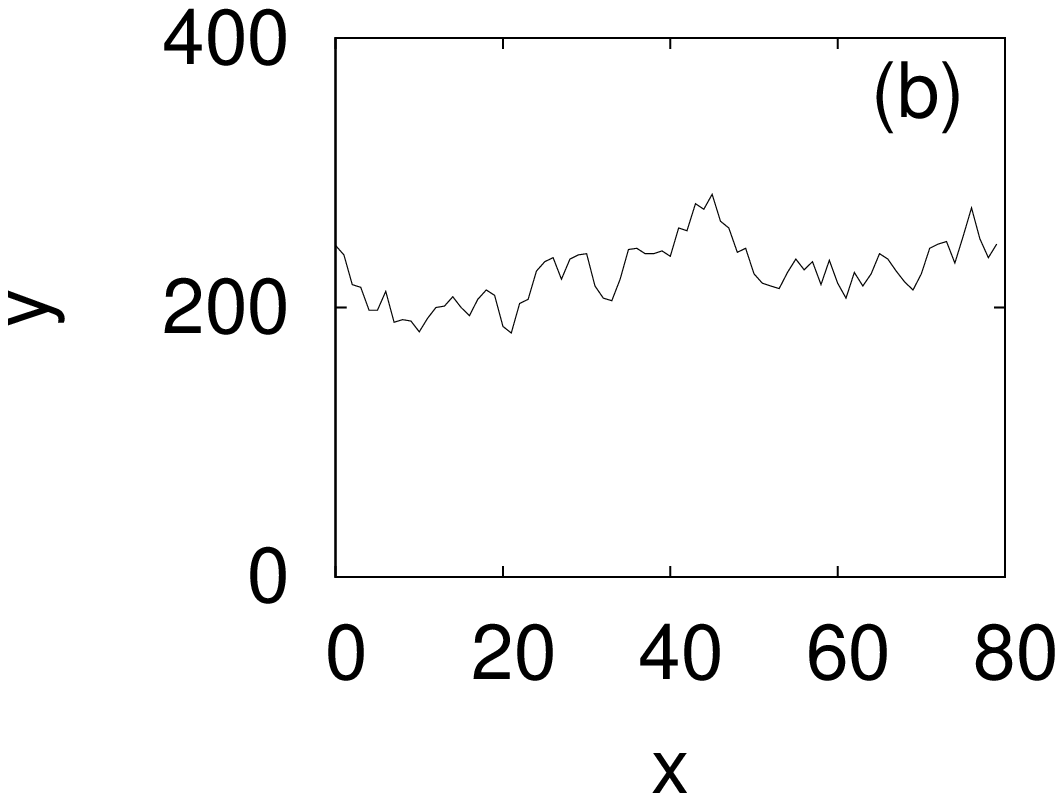}
\includegraphics[width=3.5cm]{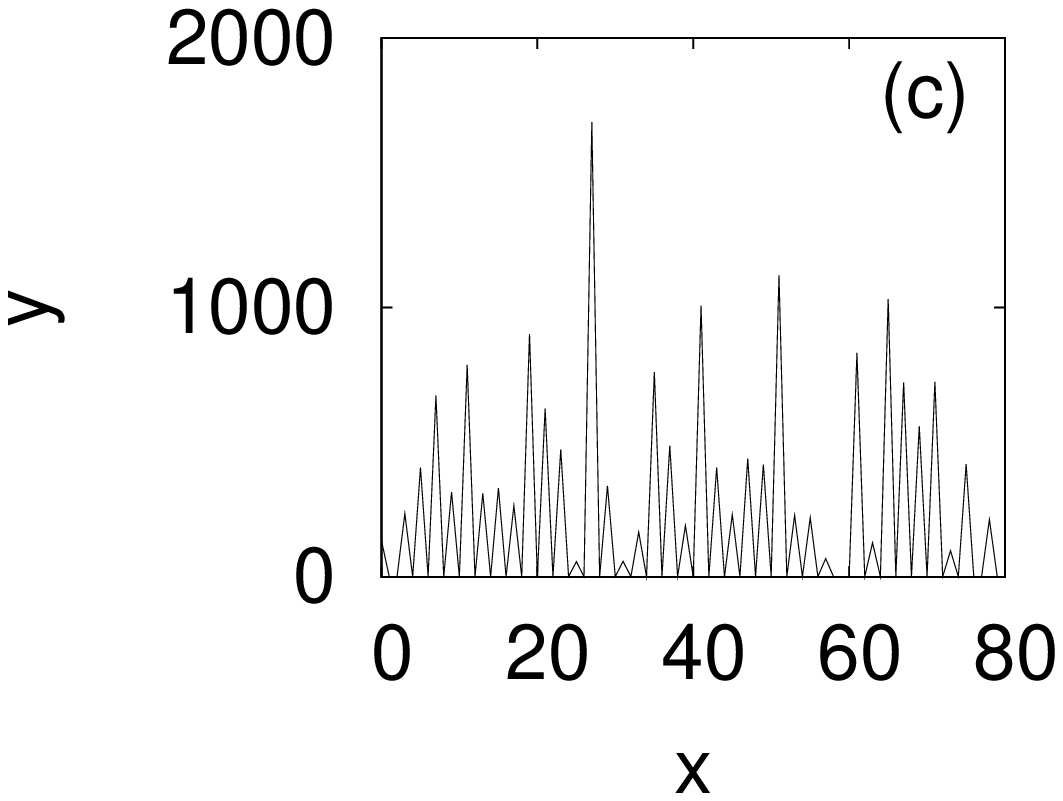}
\includegraphics[width=3.5cm]{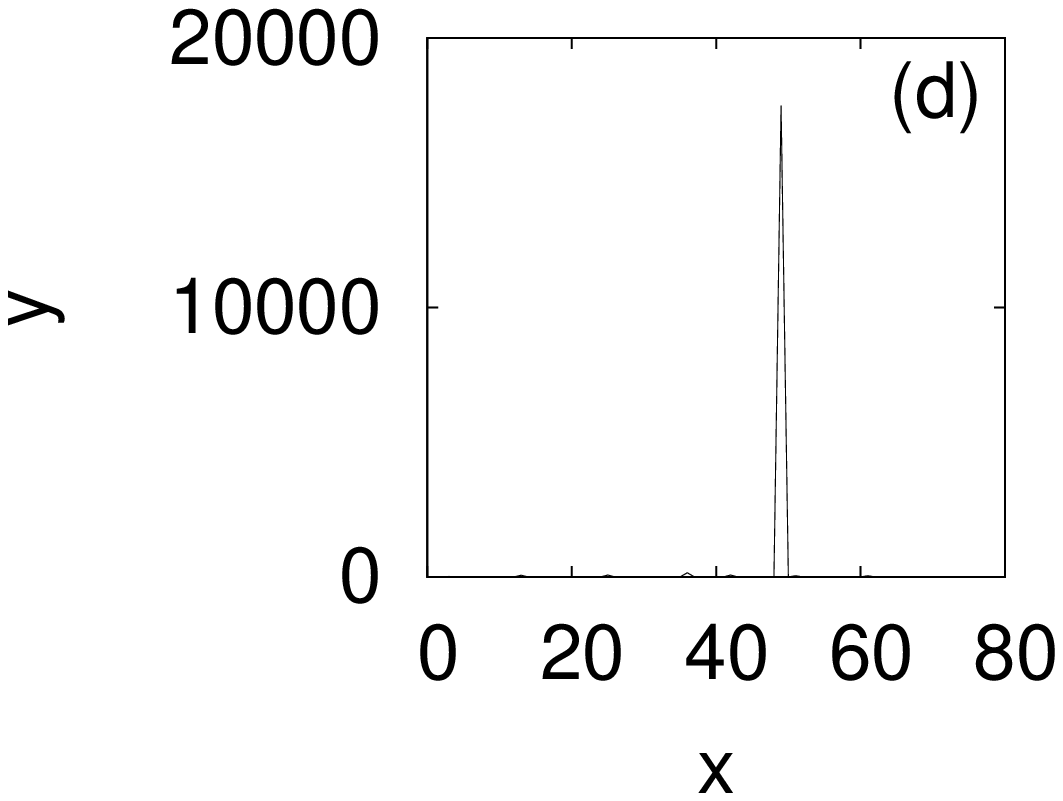}
\includegraphics[width=3.5cm]{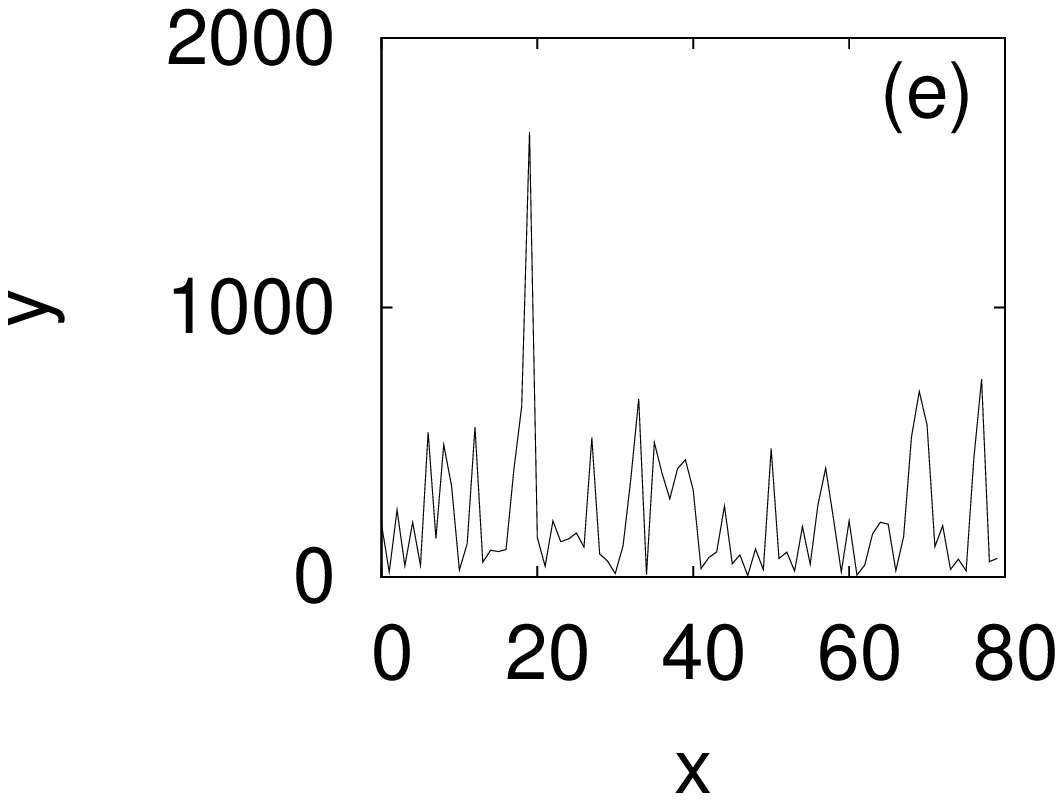}
\caption{\label{config} Typical configurations for all phases from Fig.~\ref{phases}: (a)-droplet, (b)-correlated fluid, (c)-antiferromagnetic, (d)-localised, (e)-uncorrelated fluid.}
\end{figure}

We have used the parameters $\sigma,\gamma$ and $\delta$ to determine the phase diagram shown in Fig.~\ref{phases} (see also Fig.~\ref{phases2} for examples of plots of $\sigma,\gamma,\delta$) in the phase plane of the parameters $c_1$ and $c_2$, by simulating our model for fixed $N=80, M=18100$, and different pairs of $c_1,c_2$. Snapshots of typical configurations in each phase are shown in Fig.~\ref{config}. Our phase diagram includes both positive and negative $c_1,c_2$. One might be worried that negative coupling constants should not have any physical meaning in the CDT, because the effective action $S_{\rm eff}$ would be unbounded from below for negative $c_1,c_2$ and hence the partition function was ill-defined. However, as we consider here the system with a finite number of sites $N$ and particles $M$, the action is bounded and the partition function is well defined.

Looking at Fig.~\ref{phases}, we can distinguish five different phases in the $(c_1,c_2)$ plane for fixed $N,M$:
\begin{itemize}
\item{Droplet phase: a finite fraction of particles (typically almost all particles) form a bell-shaped condensate extended over $W\gg 1$ sites of the lattice. The shape of the condensate can be approximated by Eq.~(\ref{cos3}). The droplet phase is observed for $c_1>0$ and $c_2>c_{2,\rm crit}(c_1)$, where the shape of the critical curve $c_{2,\rm crit}$ depends also on $N$ and $M$. This phase corresponds to the macroscopic universe phase ``C'' in CDT. The width $W$ and other properties of the condensate will be discussed in Section \ref{sec:droplet}. The values of the order parameters are as follows: $\sigma$ is of order $1/W$, $\gamma = 0$ and $\delta > 0$.}
\item{Correlated fluid: particles are distributed approximately uniformly over all sites of the lattice. The occupation numbers fluctuate around the average value $\left<m_i\right>=\rho$, but the typical size of fluctuations is small
as compared to the average. This phase is observed for $c_1>0$ and $c_2 < c_{2,\rm crit}(c_1)$. In the thermodynamic limit, we expect the order parameters to be $\sigma=0$ (of order $1/N$ for finite system), 
$\gamma>0$ and $\delta>0$.}
\item{Antiferromagnetic fluid: typical configurations contain alternated occupied/empty (i.e., containing only one particle) sites. 
This phase is observed when both $c_1$ and $c_2$ are negative. The number of empty sites increases when $c_1$ or $c_2$ grow. In the thermodynamic limit,
the order parameters in this phase are $\sigma=0$ (of order $2/N$ for finite $N$), $\gamma=0$ and $\delta=0$.}
\item{Localised phase: in a typical configuration, almost all particles occupy a single site, while the remaining sites have only small numbers of particles of order $O(1)$. This phase is observed for $c_1<0$ and $c_2>0$. The order parameters are $\sigma=1$, $\gamma=0$ and $\delta>0$. 
This phase may correspond to phase ``B'' in CDT.}
\item{Uncorrelated fluid: Particle occupation numbers are uncorrelated and there is no condensation regardless of the density of particles $\rho$. This phase is observed in a small region close to the origin of the $(c_1,c_2)$ plane: $c_1 \approx 0, c_2\approx 0$ and it may correspond to ``A'' of the CDT model.}
\end{itemize}
Interestingly, as we have already mentioned, there are two new phases: the correlated-fluid phase and the antiferromagnetic-fluid phase, which have not been observed in computer simulations of CDT. In next sections we shall present some arguments supporting the existence of these new phases in the full CDT quantum gravity model.

We shall now give a crude mean-field argument supporting our phase diagram, based on estimating the value of the action
\bq
S =\sum_i \left(c_1\frac{(m_{i+1}-m_i)^2}{(m_i+m_{i+1})/2} + c_2 m_i^{1/3}\right), \label{eq:ac1}
\eq
for typical configurations in different phases, and assuming that, for given $c_1$ and $c_2$, the phase with the least value of the action is selected. Although we neglect quantum fluctuations of $m_i$'s in this section, we shall see that our approach reproduces the phase diagram quite well. Quantum fluctuations will be discussed in the next section.

\begin{figure}
\centering
\psfrag{c1}{$c_1$} \psfrag{c2}{$c_2$}
\includegraphics[width=8cm]{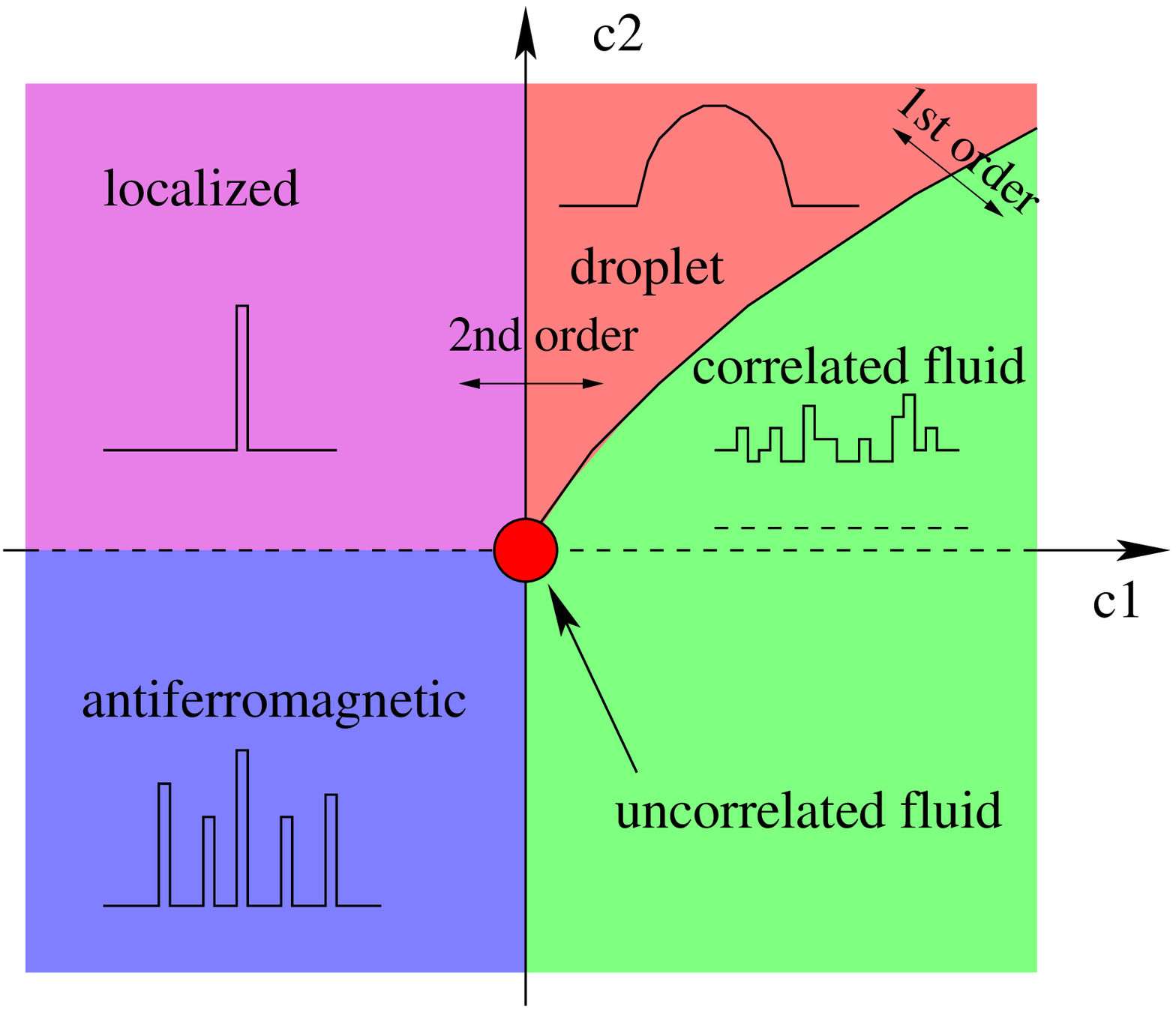}
\caption{\label{diagram_theor}Phase diagram obtained by comparing the action of typical configurations in different phases.}
\end{figure}

The mean-field action for the droplet of width $W$ shown in Fig. \ref{config}a can be approximated as
\bq
S_{\rm droplet} \approx 2c_1 \frac{M}{W} \int_0^W \frac{(h((i+1)/W)-h(i/W))^2}{h((i+1)/W)+h(i/W)} di + c_2 \left(\frac{M}{W}\right)^{1/3} \int_0^W h(i)^{1/3} di, \label{eq:drop1}
\eq
where we have assumed that the average shape of the droplet is $m_i = (M/W)h(i/W)$ and that fluctuations can be neglected in the limit of large $M$. We assume that $h(x)$ is fixed and the only degree of freedom is the width $W$ of the droplet. Equation (\ref{eq:drop1}) can be further simplified if $h((i+1)/W) \cong h(i/W) + h'(i/W)/W$,
\bq
S_{\rm droplet} \approx c_1 \frac{M}{W^2} \int_0^1 dx \frac{h'(x)^2}{h(x)} + c_2 W^{2/3} M^{1/3} \int_0^1 dx h(x)^{1/3} \ . \label{eq:drop2}
\eq
The integrals over $dx$ will be explicitly calculated later, now we just treat them as two unknown constants. Searching for $W$ which minimizes the action we obtain $W\sim (c_1/c_2)^{3/8} M^{1/4}$ and, finally,
\bq
S_{\rm droplet} \propto c_1^{1/4} c_2^{3/4} M^{1/2} \ .
\eq
We see that the above calculation predicts the extension $W$ of the condensate to grow as $\sim M^{1/4}$. We shall come back to that later. Now, let us consider the energy of the correlated fluid phase (see Fig. \ref{config}b):
\bq
S_{\rm corr. fluid} \approx N c_1 \frac{\left<(m_{i+1}-m_i)^2\right>}{\rho} + c_2 N \rho^{1/3} \ .
\eq
Assuming that $m_i = \rho + \Delta m_i$ where $\Delta m_i$ is of order $\sqrt{\rho}$ due to stochastic fluctuations, we obtain
\bq
S_{\rm corr. fluid} \propto N c_1 + c_2 N^{2/3} M^{1/3} \ .
\eq
The value of the action for a typical configuration in the antiferromagnetic phase (see Fig. \ref{config}c) is
\bq
S_{\rm antiferr.} \approx 2c_1 M + c_2 n^{2/3} M^{1/3}
\eq
where we have assumed that there are $n$ peaks of height $\approx M/n$, separated by empty sites. We can use the last formula also to estimate
the action in the localised phase (see Fig. \ref{config}d) by setting $n=1$:
\bq
S_{\rm localised} \approx 2c_1 M + c_2 M^{1/3} \ .
\eq
Comparing the values of the action for different $c_1,c_2$ and taking the least one, we obtain for large $N,M$ the phase diagram shown in Fig. \ref{diagram_theor}.
The diagram agrees qualitatively with the experimentally obtained one in Fig.~\ref{phases}. The lines separating different phases are at $c_1=0$ and $c_2=0$, except for a line between the droplet and the correlated fluid phase, which has a more complicated shape and will be discussed in Sec.~\ref{sec:corrfl}.
The reader may wonder why we did not estimate the action in the uncorrelated fluid phase. The reason is that this
phase is dominated by fluctuations (entropy) rather than by the action (energy) (\ref{eq:ac1}) which vanishes for $c_1=c_2=0$.
Although this phase exists only at a single point $(c_1,c_2)=(0,0)$ in the phase space in the thermodynamic limit, we expect that for finite systems we discuss here, the uncorrelated-fluid phase extends to a small region around $(c_1,c_2)=(0,0)$. 

We conclude this section with a technical remark. Because our model is motivated by the CDT model of quantum gravity, we prefer to
use the language of quantum physics rather than that of statistical physics in the paper. If one used statistical physics
language instead, one would replace the action $S$ by $\beta E$, where
$\beta=c_1$ would be the inverse temperature, $E$ would be the energy of configurations, and $c_2/c_1$ would be the second parameter (besides $\beta$) of our model. The partition function could then be written as $Z = e^{-\beta F} = \sum_{\{m\}} e^{-\beta E}$, 
where $F$ would correspond to the free energy of the system, including the entropic contribution 
coming from the sum over all microstates. In quantum physics, $F$ is rather referred to as the effective action and the entropic contribution as to the contribution from quantum fluctuations. 
In the next section we shall estimate the contribution from quantum fluctuations to the droplet phase
and show that these fluctuations lead to the widening of the effective universe as compared to the classical de-Sitter solution.

\section{Droplet phase - the macroscopic universe\label{sec:droplet}}
\begin{figure}
		\psfrag{w}{$W$} \psfrag{M}{$M$}
		\includegraphics[width=4cm]{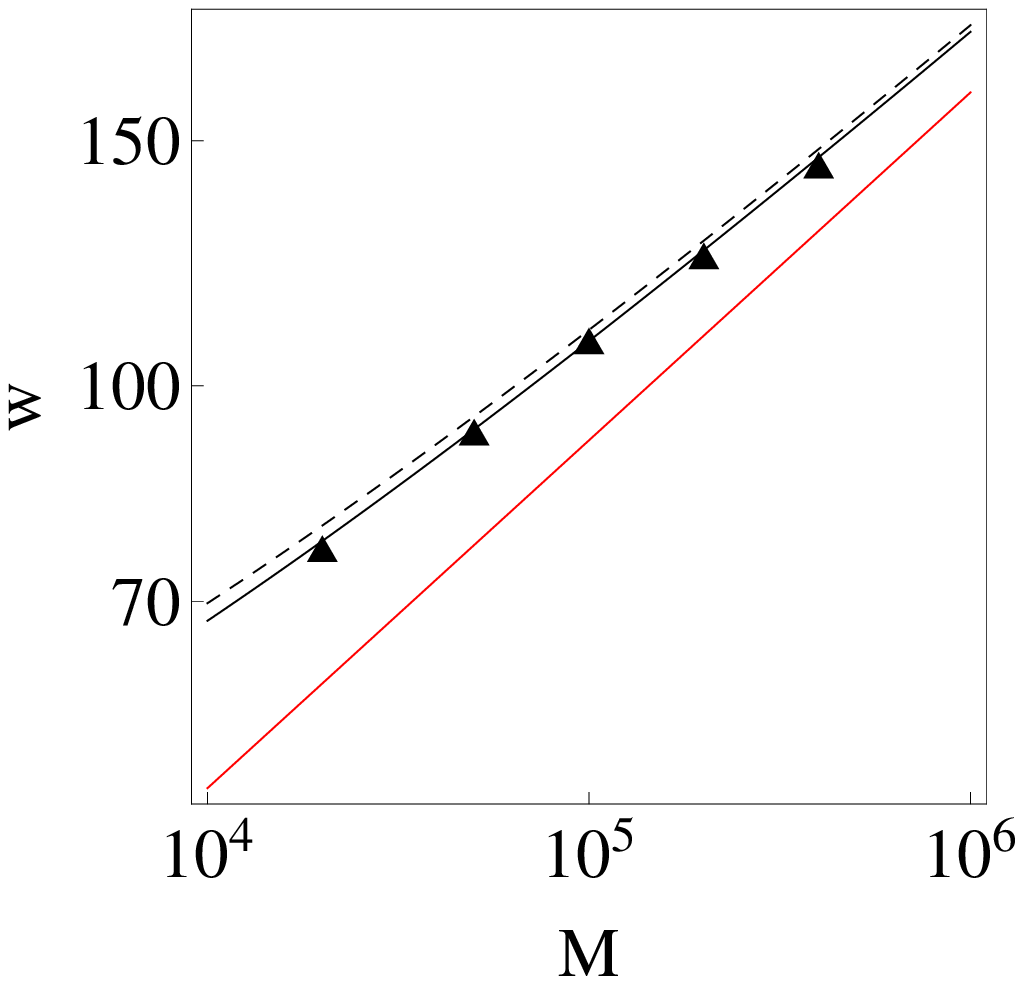}
		\includegraphics[width=4cm]{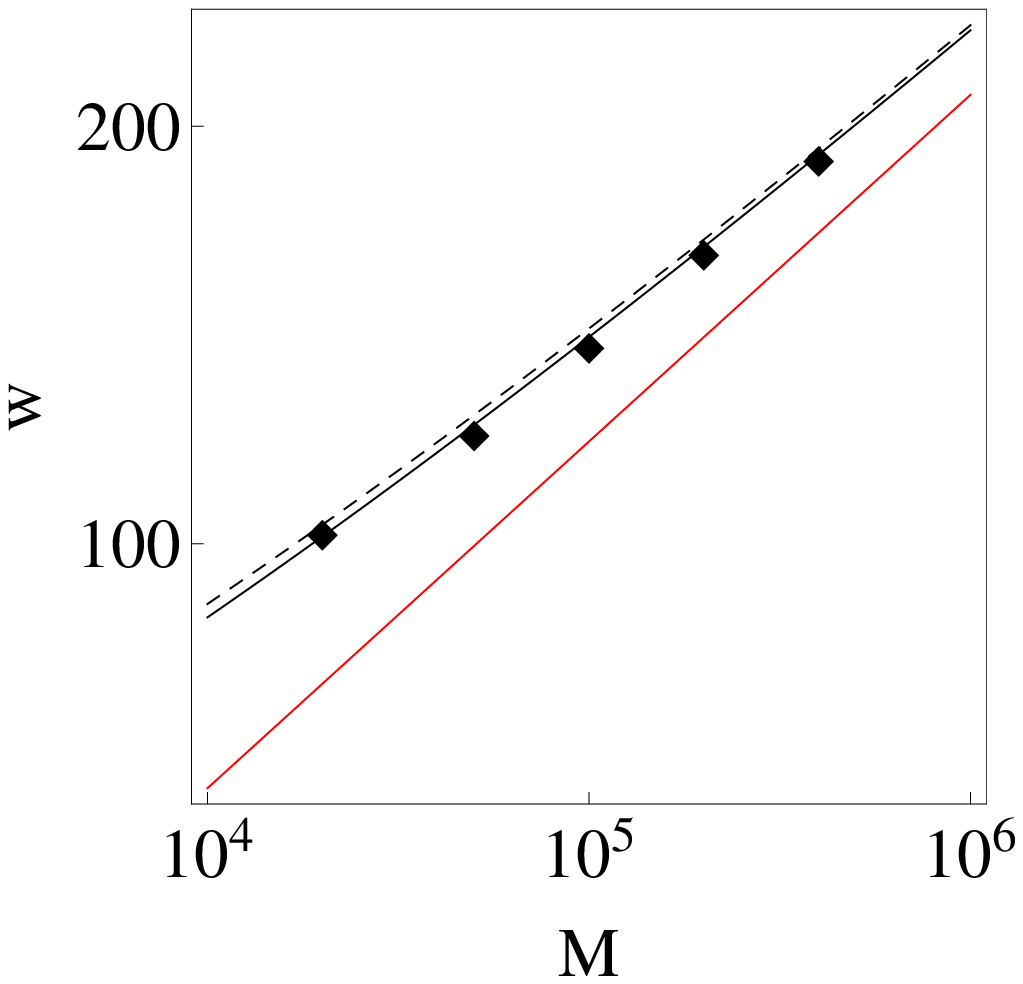}
		\includegraphics[width=3.9cm]{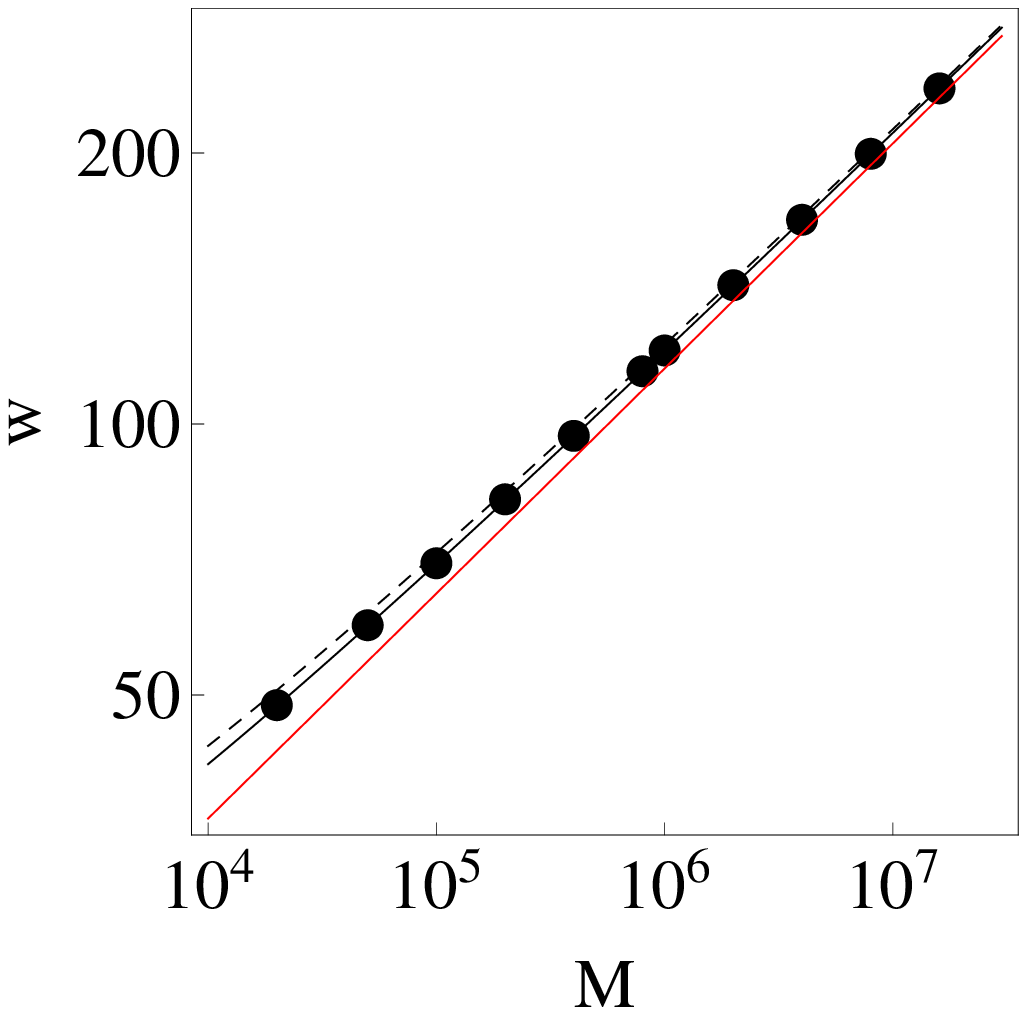}
		\includegraphics[width=3.9cm]{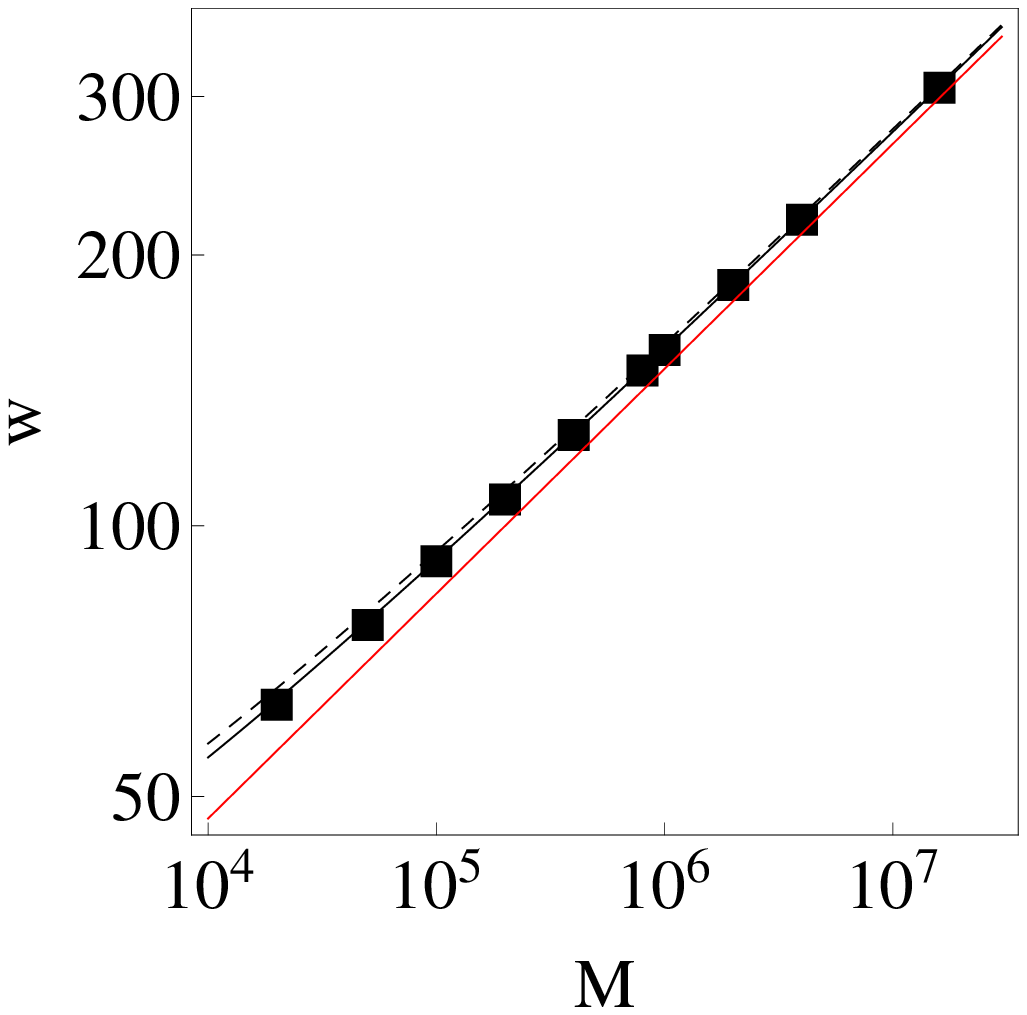}
		\caption{\label{fig:widths} The width $W$ versus the number of particles $M$ for different pairs $(c_1,c_2)=(1,2),(2,2),(1,5),(2,5)$ (from left to right). Black symbols correspond to numerical data, red line shows the classical solution $W=W_0(M)$ (\ref{w0}), black dashed line corresponds to the solution $W=W_1(M)$ of Eq.~(\ref{eq:finalw}) taking into account quantum corrections and black solid line shows the quantum solution including interface effects $W=W_2(M)$ from Eq.~(\ref{w2}).}
\end{figure}

In the droplet phase, which exists for positive coupling constants $c_1,c_2$, the condensate takes the form of an extended ``droplet''. 
\begin{figure}
\centering
	\psfrag{x}{$t$}\psfrag{y}{$m(t)$}
	\includegraphics[width=8cm]{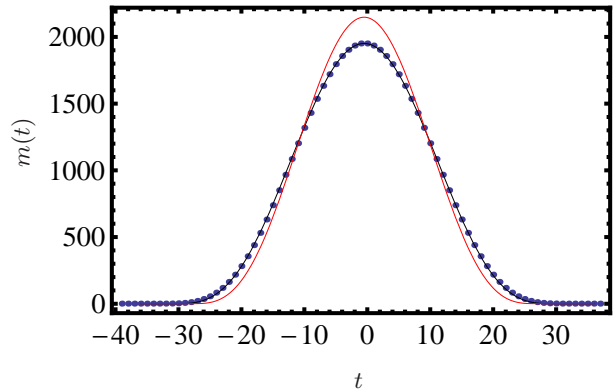}
\caption{\label{droplet_MC_th} Shape of the universe in the droplet phase
averaged over $n=1000000$ configurations for $c_1=1$ and $c_2=5$ and 
$N=256$ and $M=50000$ (blue points), compared to the $\cos^3$ shape (Eqs.~(\ref{mt}) and (\ref{hx})) with the width parameter given by the classical solution Eq.~(\ref{w0}) $W=W_0(M)=55.5$ (red curve), and a more accurate result
including quantum and finite size corrections Eq.~(\ref{w2})
$W=W_2(M)=59.64$ (black curve).}
\end{figure}
In Fig. \ref{droplet_MC_th} we show the average shape of this droplet obtained in numerical simulations (see the appendix for details). The envelope of the droplet has a $\cos^3$ form and its extension scales as $\sim M^{1/4}$ (see Fig.~\ref{fig:widths}) as determined already in the previous section. We will now find the function $h(x)$ and calculate the integrals from Eq.~(\ref{eq:drop2}) to find the coefficient in the power law $W\sim M^{1/4}$. Let us first assume that in the limit of large system sizes $N,M\to\infty$ and $\rho=\rm const$, fluctuations of $\{m_i\}$ can be neglected, so that
\bq
m_i \equiv \bar m_i,
\eq
where $\bar m_i$ denotes the average occupation number at site $i$. The shape of the condensate can be obtained by minimising the action
\bq
S(\{ \bar m_i \}) = \sum_{i=1}^N \left[c_1\frac{(\bar m_{i+1}-\bar m_i)^2}{(\bar m_i+\bar m_{i+1})/2} + c_2 \bar m_i^{1/3}\right].
\eq
Going into the continuous limit: $\bar m_i \to m(t)$ and $\bar m_{i+1}-\bar m_i \to m'(t)$, with $m(t)$ defined on the interval $0\leq t\leq N$, we see that the following functional has to be minimised with respect to $m(t)$:
\bq
S[m(t)] = \int_{0}^{N} \left[ c_1 \frac{m'(t)^2}{m(t)}+c_2 m(t)^\frac{1}{3} \right] dt,
\eq
with an additional constraint that $\int_0^N m(t) dt = M$. Using the method of Lagrange multipliers we obtain the following Euler-Lagrange differential equation for $m(t)$:
\bq
	\frac{c_2}{3}m(t)^{-\frac{2}{3}}+c_1 \left(\frac{m'(t)}{m(t)}\right)^2-2 c_1 \frac{m''(t)}{m(t)} - a = 0.
\eq
where $a$ is the Lagrange multiplier used to fix the total number of particles $M$. 
This equation is exactly soluble:
\bq
	m(t) = \frac{M}{W} h(t/W), \label{mt}
\eq
where $h(x)$ is the ``$\cos^3$'' shape of the droplet,
\bq
	h(x) = \left\{ \begin{array}{ll} \frac{3\pi}{4} \cos^3 (\pi (x-1/2)) = \frac{3\pi}{4} \sin^3 (\pi x), & 0 <x< 1 \\ 0, & x<0\;\mbox{or}\; x>1 \end{array} \right. \label{hx} \\
\eq
and $W$ is the width of the droplet,
\bq
 W = W_0(M) = M^{1/4} \frac{3 \pi}{\sqrt{2}} \left(\frac{c_1}{c_2}\right)^{3/8} \approx 6.66432 \times M^{1/4} \left(\frac{c_1}{c_2}\right)^{3/8} .\label{w0}
\eq
Equations (\ref{mt}) and (\ref{hx}) are equivalent to Eq.~(\ref{cos3}) up the position of the centre of mass which is shifted from $x=0$ to $x=1/2$ (we have used the freedom of shifting the droplet to $x=1/2$ for the future convenience). The width $W$ is uniquely determined by $M,c_1,c_2$ and it grows as expected as $\sim M^{1/4}$ for large systems. Equation~(\ref{mt}) shows that the average height of the droplet $M/W$ scales as $\sim M^{3/4}$. Remembering that $M$ plays the role of four-volume of the corresponding CDT model, we see that the height is proportional to the three-volume of spatial slices. This is one of the reasons why the droplet is considered to be a manifestation of a macroscopic universe in Refs. \cite{quantum1,quantum2,quantum3,quantum4}. 

The shape observed numerically closely follows the classical solution (\ref{hx}), see the red curve in Fig. \ref{droplet_MC_th}. However, the width of the droplet $W$ observed in numerical simulations is larger than the one calculated from Eq.~(\ref{w0}), as shown in Fig.~\ref{fig:widths} (red curves). The reason is that calculation that lead to Eq.~(\ref{w0}) neglect quantum fluctuations. 
We will now calculate quantum corrections to $W$ assuming that they leave the shape of the droplet intact. This assumption, as we have mentioned, is corroborated by simulations. Our reasoning follows in part the lines of Ref.~\cite{bw1} in which the spatial extension of the condensate has been calculated analytically by splitting the system into two parts: the condensate and the fluid background. Proceeding in a similar way, we assume that the total free energy $F(W)$ of the system having a condensate of width $W$ can be approximated by
\ba
F(W) \approx F_{\rm background}(N-W) + F_{\rm droplet}(W) = \ln Z_{\rm crit}(N-W,\rho_c(N-W)) + \ln Z_{\rm droplet}(W,\tilde{M}) \nonumber \\
= -W \ln \lambda_{\rm max} + \ln Z_{\rm droplet}(W,\tilde{M}) + O(N), \label{eq:pw}
\ea
where $Z_{\rm crit}$ is the canonical partition function for the system with $N-W$ sites being at the critical density, and $Z_{\rm droplet}$ is the partition function for the condensate extended over $W$ sites and containing $\tilde{M} = \left(1 - [1\!-\!W/N] \rho_{\rm crit}/\rho \right) M$ particles. If the density is high (the case relevant for CDT),  $\rho_{\rm crit}/\rho \ll 1$ and we can assume $\tilde{M} \approx M$. Equation (\ref{eq:pw}) states that the free energy of the system is the sum of free energies of the fluid and the droplet, and neglects contributions from the boundaries between these two coexisting states. The partition function for the bulk reads $Z_{\rm crit}(N-W)\sim \lambda_{\rm max}^{N-W}$, where $\lambda_{\rm max}$ is the maximal eigenvalue of the matrix $T_{mn}(z_c)$ defined in Eq.~(\ref{eq:tmn}). The partition function of the condensate $Z_{\rm droplet}(W)$ reads:
\bq
Z_{\rm droplet}(W,M) = \sum_{m_1=1}^\infty \cdots \sum_{m_W=1}^\infty \exp\left[-S_W(\{m_i\})\right] \delta\left(M-\sum_i m_i\right),
\label{Zdroplet}
\eq
where $m_0=m_{W+1}=1$ and
\bq
S_W[\{m_i\}] = \sum_{i=0}^{W} \left(2c_1\frac{(m_{i+1}-m_i)^2}{m_i+m_{i+1}} + c_2 m_i^{1/3}\right), \label{lnp2}
\eq
is the action for the droplet of size $W$. 
The standard way of estimating the contribution of fluctuations is to
expand each $m_i$ around its average value $\bar{m}_i$, $m_i = \bar{m}_i + \Delta m_i$,
and to assume that the fluctuations $\Delta m_i$ are Gaussian. In this approximation,
\bq
Z_{\rm droplet}(W) \cong e^{-S_W[\{\bar{m}_i\}]} 
\int \exp\left[ -\frac{1}{2} 
\sum_{ij} \Delta m_i \bar{A}_{ij} \Delta m_j \right]
\delta\left(\sum_i \Delta m_i\right) \prod_i d \Delta m_i ,
\label{Zd}
\eq
where $\Delta m_i$ are now continuous variables and the matrix $\bar{A}$ is the matrix of second derivatives (the Hessian),
\bq
\bar{A}_{ij} = \frac{\partial^2 S_W}{\partial m_i \partial m_j}\bigg|_{m_i=\bar{m}_i} ,
\eq
calculated for $\{m_i\}$ which correspond to the classical solution $\bar{m}_i=M/W h(i/W)$ (see Eqs.~(\ref{mt}) and (\ref{hx})). Using the integral representation of the Dirac delta
\bq
\delta(k) = \int^{\infty}_{-\infty} \frac{dq}{2\pi} e^{iqk} ,
\label{deltak}
\eq
we obtain
\bq
Z_{\rm droplet}(W) \cong e^{-S_W[\{\bar{m}_i\}]} 
\int_{-\infty}^{+\infty} \frac{dq}{2\pi}
\int  \exp\left[ -\frac{1}{2} 
\sum_{ij} \Delta m_i \bar{A}_{ij} \Delta m_j + i q \sum_i m_i \right] \prod_i d \Delta m_i .
\label{Zd2}
\eq
We can now calculate the Gaussian integral over $\Delta m_i$'s using the standard result:
\bq
	\int d^W n \exp\left[-\frac{1}{2} \sum_{i,j} \bar{A}_{ij} n_i n_j + \sum_j n_j b_j\right] = \sqrt{\frac{(2\pi)^W}{\det \bar A}}\exp\left[\frac{1}{2} \sum_{i,j} b_i b_j (\bar{A}^{-1})_{ij} \right],
\eq
where $\bar{A}^{-1}$ denotes the inverse of $\bar{A}$. Taking $b_j = i q$ for all $j$ we have
\bq
	Z_{\rm droplet}(W) \cong e^{-S_W[\{\bar{m}_i\}]} \sqrt{\frac{(2\pi)^W}{\det \bar A}} \int_{-\infty}^{+\infty} \frac{dq}{2\pi} \exp\left[-\frac{1}{2}q^2 \sum_{i,j} (\bar{A}^{-1})_{ij} \right],
	\label{Zd3}
\eq
and, performing the last Gaussian integral over $q$, we obtain that
\bq
	F_{\rm droplet}(W) = \ln Z_{\rm droplet}(W) \cong - S_W[\{\bar{m}_i\}] + Q(W), \label{Fdrop}
\eq
where $Q(W)$ correspond to a quantum correction to the free energy:
\bq
	Q(W) = \frac{W}{2} \ln(2\pi) - \frac{1}{2}\ln \det \bar{A} - \frac{1}{2} \ln \left( \sum_{i,j} (\bar{A}^{-1})_{ij} \right). \label{eq:quant}
\eq
The first term in Eq.~(\ref{Fdrop}) is just the action (\ref{lnp2}) calculated along the classical trajectory and it can be easily evaluated in the continuous approximation:
\bq
	S_W[\{\bar{m}_i\}] \cong \int_0^W \left(c_1\frac{(m'(t))^2}{m(t)} + c_2 m(t)^{1/3}\right) = \frac{9\pi^2 c_1 M}{2W^2} + \frac{6^{1/3} c_2 M^{1/3} W^{2/3}}{\pi^{2/3}}, \label{SWfinal}
\eq
where we have inserted $m(t)$ from Eqs.~(\ref{mt}) and (\ref{hx}). The quantum contribution $Q(W)$ to the effective action from Eq.~(\ref{eq:quant}) consists of three terms. The first term $ \frac{W}{2} \ln(2\pi)$ is trivial. The second term $- \frac{1}{2}\ln \det \bar{A}$ is more complicated because it contains the determinant of $\bar A$. To evaluate this determinant, we first observe that the matrix $\bar A$ is tridiagonal, with only non-zero elements being
\ba
	\bar{A}_{ii} &=& -\frac{2 c_2}{9 \bar{m}_i^{5/3}}+\frac{4 c_1 (-\bar{m}_{i-1}+\bar{m}_i)^2}{(\bar{m}_{i-1}+\bar{m}_i)^3}-\frac{8 c_1 (-\bar{m}_{i-1}+\bar{m}_i)}{(\bar{m}_{i-1}+\bar{m}_i)^2}+\frac{4 c_1}{\bar{m}_{i-1}+\bar{m}_i} \nonumber \\
	& &+ \frac{4 c_1 (-\bar{m}_i+\bar{m}_{i+1})^2}{(\bar{m}_i+\bar{m}_{i+1})^3}+\frac{8 c_1 (-\bar{m}_i+\bar{m}_{i+1})}{(\bar{m}_i+\bar{m}_{i+1})^2}+\frac{4 c_1}{\bar{m}_i+\bar{m}_{i+1}} \cong \frac{4c_1 W}{M h(i/W)}, \\
	\bar{A}_{i,i\pm 1} &=& \frac{4 c_1 (-\bar{m}_i+\bar{m}_{i\pm 1})^2}{(\bar{m}_i+\bar{m}_{i\pm 1})^3}-\frac{4 c_1}{\bar{m}_i+\bar{m}_{i\pm 1}} \cong -\frac{2c_1 W}{M h(i/W)}.
\ea
We see that $\bar{A}_{i,i\pm 1} \approx -\frac{1}{2} \bar{A}_{ii}$
so the determinant $\det \bar A$ can be approximated by 
$\det \bar A \approx (\det \bar a) \prod_{i=1}^W \bar{A}_{ii}$,
where the matrix $\bar a$ is a tridiagonal matrix 
with diagonal elements $\bar{a}_{ii}=1$ and off-diagonal ones $\bar{a}_{i,i\pm 1}=-1/2$. One should note that due to the periodic boundary conditions also the corner elements $\bar{a}_{1N}$ and $\bar{a}_{N1}$ of this matrix should be in principle equal to $-1/2$. In this case the matrix $\bar{a}$ would have a zero mode. The zero mode has been however removed by fixing the position of the centre of mass to be at $N/2$. With this choice one can safely set $\bar{a}_{1N}=\bar{a}_{N1}=0$ leaving only the tridiagonal structure of the matrix $\bar{a}$. 
The determinant of this matrix $\det \bar{a} = (N+1) 2^{-N}$ is independent of $W$, hence the whole dependence of quantum corrections on $W$ is in the factor $\prod_{i=1}^W \bar{A}_{ii}$. We can now estimate that 
\bq
	\ln \det \bar{A} 
	\cong \sum_{i=1}^W \ln \frac{4c_1 W}{M h(i/W)} +O(N) \cong W \int_0^1 \ln \frac{4c_1 W}{M h(x)} dx +O(N) = W \ln \frac{128c_1 W}{3\pi M} +O(N).
\eq
This is the leading term in $Q(W)$. We shall now argue that the last term $\sum_{i,j} (\bar{A}^{-1})_{ij}$ in the quantum correction $Q(W)$ can be neglected. The reason is that because $\bar{A}_{ij} \propto W/(M h(i/W))$, elements of the inverse matrix $\bar{A}^{-1}$ have to be proportional to a product of different powers of $W,M$. Therefore, the sum $\sum_{i,j}^W (\bar{A}^{-1})_{ij}$ will also be proportional to a certain power of $M$ times a certain power of $W$ (one can show using the fact that $\bar{A}_{ij}$ is a Laplacian matrix times a diagonal matrix with elements $\sim 1/h(i/W)$ that $\sum_{i,j}^W (\bar{A}^{-1})_{ij} = M W^2 O(1)$), and its logarithm will give only a sub-leading correction $\sim \ln W$ to $Q(W)$, whose leading behaviour is $\sim W\ln W$.

\begin{figure}
	\centering
	\psfrag{M}{$M$} \psfrag{ratio}{\hspace{-1cm}$(W_{th}-W_{exp})/W_{exp}$}
	\includegraphics[width=8cm]{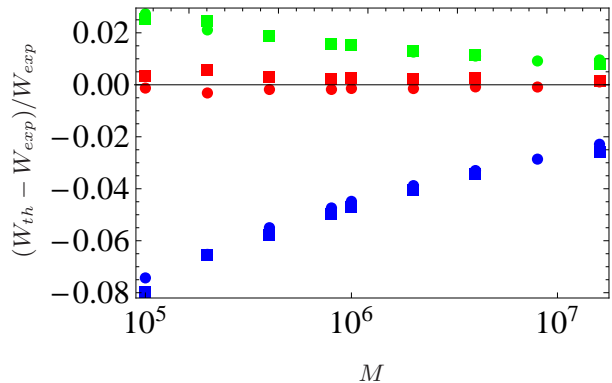}
	\caption{\label{fig:conv} Plots of a normalised deviation between theoretical $W_{th}$ and experimental $W_{exp}$ results  versus $M$. Experimental results were obtained by MC simulations of the model.  Blue symbols show the ratio $(W_{th}-W_{exp})/W_{exp}$ for classical prediction $W_{th}=W_0(M)$  from Eq.~(\ref{w0}). Green and red symbols show the ratio for quantum predictions $W_{th}=W_1(M)$ (calculated from Eq.~(\ref{eq:finalw})) and $W_{th}=W_2(M)$ (Eq.~(\ref{w2})), correspondingly	without and with interface corrections. Circles correspond to $(c_1,c_2)=(1,5)$, squares to $(c_1,c_2)=(2,5)$. One can see that the expression $W=W_2(M)$ almost ideally reproduces results of simulations.}
\end{figure}

In summary, the quantum correction approximately reads 
\bq
	Q(W) \cong \frac{W}{2}\left( \ln \frac{3\pi^2}{64c_1} - \ln W + \ln M\right) +O(N), \label{Qfinal}
\eq
and, inserting Eqs.~(\ref{Qfinal}) and (\ref{SWfinal}) into Eq.~(\ref{Fdrop}), and then Eq.~(\ref{Fdrop}) into Eq.~(\ref{eq:pw}) we obtain the final expression for the free energy of the system:
\bq
	F(W) \cong -W\ln \lambda_{\rm max} - \frac{9\pi^2 c_1 M}{2W^2} - \frac{6^{1/3} c_2 M^{1/3} W^{2/3}}{\pi^{2/3}} + \frac{W}{2}\left( \ln \frac{3\pi^2}{64c_1} - \ln W + \ln M\right) + O(N).
\eq
The width $W$ of the droplet is determined by the maximum of $F(W)$. Taking a derivative with respect to $W$ we finally arrive at an equation for the spatial extension $W$:
\bq
-\ln \lambda_{\rm max}+c_1 9\pi^2 \frac{M}{W^3}- c_2 \frac{2\cdot 6^{1/3} M^{1/3}}{3\pi^{2/3} W^{1/3}}  + \frac{1}{2} \left(\ln \frac{3 \pi ^2}{64 c_1}+\ln M-\ln W -1\right) =0. \label{eq:finalw}
\eq
In the limit of large $M$, this equation leads to the same expression as Eq.~(\ref{w0}). For finite $M$ we solve it numerically for $W$. The solution
gives a function $W=W_1(M)$ which includes quantum corrections. The maximal eigenvalue $\lambda_{\rm max}$ of the matrix $T_{mn}$ from Eq.~(\ref{eq:tmn}) which is necessary to solve Eq.~(\ref{eq:finalw}) can be determined by numerical diagonalisation of $T_{mn}$ truncated at $m,n\approx 50$. In Fig. \ref{fig:widths} we compare $W=W_1(M)$ calculated as a root of Eq.~(\ref{eq:finalw}) and $W=W_0(M)$ 
obtained from the classical formula~(\ref{w0}). In the same plot we also show values of $W$ measured in simulations of the model for different $c_1,c_2$. We see that the solution $W=W_1(M)$ which takes into account quantum corrections reproduces the data much better than the classical solution 
$W=W_0(M)$ from Eq.~(\ref{w0}). The agreement could be further improved by taking into account interactions on the interface between the droplet and the fluid, where the fluctuations $\Delta m_i$ become non-Gaussian. We will not do this here but instead we observe that subtracting a small correction from $W_1(M)$, 
\bq
	W_2(M) = W_1(M) - 2, \label{w2}
\eq
is enough to almost perfectly reproduce the data as shown in Figs.~\ref{fig:widths}
and \ref{fig:conv}. A physical meaning of this correction could be that 
interactions at the interface droplet-fluid exert a pressure 
on the droplet that shifts its boundaries towards the centre of mass 
by one lattice unit on each side of the droplet. 

\section{Correlated-fluid phase\label{sec:corrfl}}
In models such as B-in-B \cite{b-in-b} or ZRP \cite{evans0501338} one usually fixes the density $\rho=M/N$ of particles and takes the thermodynamic limit $M,N\to\infty$. The condensate emerges in this limit above the critical density $\rho_c$. The same remains true in our model. However, there is another important limit here, namely $M^{1/4}/N \equiv w = \rm const$ and $M,N\to\infty$. In this limit, the width $W\sim M^{1/4}$ of the condensate becomes a finite fraction of the system size $N$.

It turns out that there is a new phase transition as a function of the parameter $w$: when the width of the condensate becomes equal to $N$, both borders of the condensate merge together. The envelope of the condensate loses its $\cos^3$ shape and becomes flat: mean occupation numbers $\bar m_i = M/N \propto N^3$ are much larger than 1, and fluctuations which are of order $\sqrt{M/N}$ are not powerful enough to cause $\{m_i\}$ to drop to $m_i \approx 1$. Therefore, the condensate no longer separates from the background. We shall stress that the existence of this phase is possible only due to periodic boundary conditions. If boundary conditions were fixed, i.e., $m_1=m_N=\rm const$, the droplet would not disappear but only changed its shape for $W> N$.

\begin{figure}
	\psfrag{x}{$k$} \psfrag{y}{$A(k)$}
	\includegraphics[width=6cm]{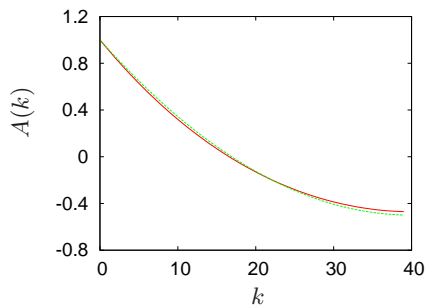}
	\caption{\label{fig:Akcf}Plot of $A(k)$ in the correlated-fluid phase (solid red line) calculated from Eq.~(\ref{eq:akfinal}) and compared to Monte-Carlo simulations for $M=72400, N=80, c_1=1.0, c_2=-0.5$ (dashed green line).}
\end{figure}

The correlated-fluid phase is not the same as the weakly-correlated fluid phase below $\rho_c$. In particular, correlations between different $m_i$'s are very strong in this phase. To calculate correlations ${\rm cov}(m_j,m_k) = \overline{m_j m_k}-\bar m_j\bar m_k$, let us first observe that the partition function (\ref{zdef}) can be approximated in this phase as
\bq
	Z_{\rm corr. fluid}(N,M) \approx \sum_{m_1=-\infty}^\infty ... \sum_{m_N=-\infty}^\infty \exp\left[ - \sum_i \left(c_1\frac{(m_{i+1}-m_i)^2}{\rho} + c_2 \rho^{1/3}\right)\right] \delta\left(\sum_i m_i-M\right), \label{zcf}	
\eq
because the average occupation numbers $\bar m_i \approx \rho$ and, since we anticipate that $\sqrt{{\rm var}(m_i)} \ll \bar m_i$, we can focus on small deviations only. If we now replace the sum by an $N$-dimensional integral over $m_1,\dots,m_N$, Eq.~(\ref{zcf}) reduces to a Gaussian integral with the constraint on the total number of particles. We can subsequently get rid of the Dirac delta by replacing it by
\bq
	\delta\left(x\right) = \lim_{\sigma\to 0} \frac{1}{\sqrt{2\pi}\sigma} e^{-\frac{x^2}{2\sigma^2}} \ .
\eq
We now define an auxiliary function $G(M,N,\vec{u})$ with auxiliary variables $\vec{u}$:
\bq
	G(M,N,\vec{u}) = \lim_{\sigma\to 0} \int {\rm d}^N m \frac{1}{\sqrt{2\pi}\sigma} \exp\left[-\frac{M^2}{2\sigma^2} -\frac{1}{2} \vec{m}^T A\vec{m} + (\vec{b} + \vec{u})^T \vec{m} \right], \label{gmnucf}
\eq
in which $b_i = M/\sigma^2$ and $A_{ij} = -\frac{2c_1}{\rho} \Delta_{ij} + \frac{1}{\sigma^2} \delta_{ij}$, where $\Delta_{ij}$ denotes a 1d discrete Laplacian with periodic boundary conditions, and $\delta_{ij}$ is the Kronecker delta. We have: 
\bq
	{\rm cov}(m_j,m_k) = \overline{m_j m_k} - \bar m_j \bar m_k = \left[\frac{{\rm d}}{{\rm d}u_j} \frac{{\rm d}}{{\rm d}u_k} \ln G(M,N,\vec{u}) \right]_{\vec{u}=0}. \label{mjmk}
\eq
The Gaussian integral in Eq.~(\ref{gmnucf}) can be performed exactly:
\bq
	G(M,N,\vec{u}) = \lim_{\sigma\to 0} \sqrt{\frac{(2\pi)^{N-1}}{\sigma^2 \det A}} \exp\left[ \vec{b}^T A^{-1} \vec{u} + \frac{1}{2} \vec{u}^T A^{-1} \vec{u}\right],
\eq
and we obtain that
\ba
	{\rm var}(m_j) &=& {\rm cov}(m_j,m_j) = \lim_{\sigma\to 0} \left(A^{-1}\right)_{jj}, \\
	{\rm cov}(m_j,m_k) &=& \lim_{\sigma\to 0} \left(A^{-1}\right)_{jk}.
\ea
The inverse matrix $A^{-1}$ which appears in these formulas can be calculated using spectral decomposition of the matrix $A$:
\ba
	A_{jk} &=& \sum_i \lambda_i \psi_{i,j} \psi_{i,k}, \\
	A^{-1}_{jk} &=& \sum_i \lambda_i^{-1} \psi_{i,j} \psi_{i,k}, \label{ajk1}
\ea
in which $\{\lambda_i\}$ and $\{\vec{\psi}_{i}\}$ are the eigenvalues and the corresponding normalised eigenvectors of $A$, respectively, 
\bq
	\lambda_k = \left\{\begin{array}{ll} N/\sigma^2 & k=1\\ 8c_1/\rho & k=2\\ \frac{8c_1}{\rho} \sin^2(\pi (k-1)/2N) & k=3,5,7 \dots\\
	\frac{8c_1}{\rho} \sin^2(\pi (k-2)/2N) & k=4,6,8 \dots\\
	\end{array}\right.
	\quad (\vec{\psi_{k}})_j = \left\{\begin{array}{ll} 1/\sqrt{N} & k=1\\ (-1)^j/\sqrt{N} & k=2\\ \frac{\cos(\pi j(k-1)/N)}{\sqrt{N/2}} & k=3,5,7,\dots\\ \frac{\sin(\pi j(k-2)/N)}{\sqrt{N/2}} & k=4,6,8,\dots\end{array}\right.
\eq
Using the expansion (\ref{ajk1}) and taking the limit $\sigma\to 0$
we obtain for large $N$:
\ba
	{\rm var}(m_j) &=& M/(24c_1),\\
	{\rm cov}(m_j,m_k) &=& \frac{M}{4c_1\pi^2} \sum_{n=1}^\infty \frac{1}{n^2} \cos\left(\frac{2\pi n (k-j)}{N}\right) \ .
\ea
This means that the correlation function $A(k)={\rm cov}(m_1,m_{k+1})/{\rm var}(m_1)$ behaves as
\bq
	A(k) = \frac{6}{\pi^2} \sum_{n=1}^\infty \frac{1}{n^2} \cos\left(\frac{2\pi n k}{N}\right), \label{eq:akfinal}
\eq
and does not depend either on $c_1,c_2$, or the number of particles. In Fig. \ref{fig:Akcf} we show that $A(k)$ calculated from the above equation agrees very well with the result of numerical simulations.

\begin{figure}
\centering
\psfrag{x}{$c_1 M^{2/3} N^{-8/3}$} \psfrag{y}{$c_2$}
\includegraphics[width=11cm]{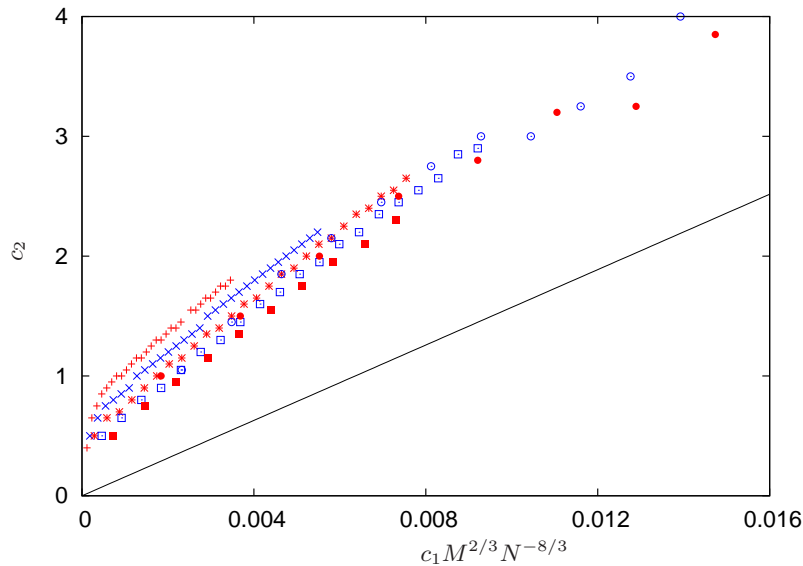}
\caption{\label{scalf}Rescaled phase line between the droplet phase and the correlated-fluid phase for systems with different number of particles $M=4525,9050,18100,\dots,289600$, and $N=80$. Solid black line is our theoretical prediction (\ref{eq:c2_as_c1}). }
\end{figure}

We shall now discuss the phase transition between the droplet and the correlated fluid phase. For any fixed $w=M^{1/4}/N$, there is a critical line in the $(c_1,c_2)$ phase plane which separates these phases. In the limit of large $N,M$ and fixed $w$, the line can be determined from the condition that $W(M)=N$ (Eq.~(\ref{w0})):
\bq
	c_{2,\rm crit}(c_1) \approx \frac{M^{2/3}}{N^{8/3}} r^{8/3} c_1, \label{eq:c2_as_c1}
\eq
where $r$ is the proportionality coefficient $r\approx 6.66432$ from Eq.~(\ref{w0}). 
This means that if we plot the transition lines determined in computer simulations for different $M,N$, and rescale $c_1\to \frac{M^{2/3}}{N^{8/3}} c_1$, all of them should collapse onto a single line. We show in Fig.~\ref{scalf} that such a collapse indeed seems to take place for large system sizes. However, for the largest $M=289600$ for which we were able to obtain the phase diagram numerically, the data points are still quite far from the theoretical line. We believe that this is caused by a very slow convergence towards the asymptotic result (\ref{eq:c2_as_c1}) due to finite-size corrections which are very strong in the region between the droplet and the fluid.

\begin{figure}
\centering
\psfrag{x}{$\gamma$} \psfrag{y}{$P(\gamma)$}
\includegraphics[width=6cm]{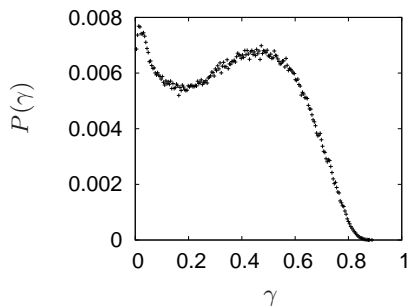}
\caption{\label{orddis} Probability of order parameter $\gamma$ for droplet-correlated fluid phase transition, $N=80$, $M=18100$, $c1=0.5$, $c2=1.29$.}
\end{figure}

The phase transition is of the first order. One can see this by observing that the two phases coexist at the transition point with a characteristic binomial structure of the distribution of the order parameter. The double maximum seen in Fig.~\ref{orddis} indicates that the system jumps from 
one phase to another. This is a typical feature of the 1st-order transition.

\section{Other phases}
\begin{figure}
\psfrag{m}{$m$}\psfrag{pm}{$p_{\rm cond}(m)$}
\includegraphics[width=8cm]{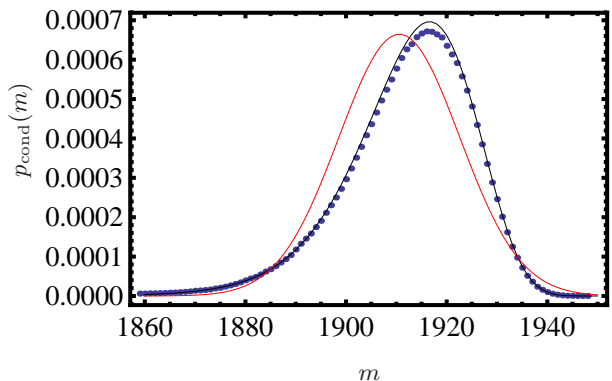}
\caption{\label{fig:pcondlf}Probability $p_{\rm cond}(m)$ that the localised condensate has $m$ particles. Blue symbols: computer simulations for $N=50,M=2000,c_1=0,c_2=5$. Black line: exact probability distribution obtained from $Z(N,M)$ calculated recursively as in Ref.~\cite{wbbj}. Red line: approximate formula (\ref{pcondapprox}) normalised so that $\sum_m p_{\rm cond} = 1/N$.}
\end{figure}

We shall now briefly discuss three other phases which appear in our model: localised (condensed) phase, antiferromagnetic fluid, and uncorrelated fluid.
For positive $c_1,c_2$, the width $W$ of the droplet decreases with decreasing $c_1$ as seen from Eq.~(\ref{w0}). Finally, at the point of $c_1=0$, the width formally reaches zero. This means that the condensate becomes localised at a single site. For $c_1=0$, the partition function reads
\bq
	Z(N,M) = \sum_{m_1=1}^M ... \sum_{m_N=1}^M \exp\left[ - c_2 \sum_i m_i^{1/3}\right] \delta\left(\sum_i m_i-M\right),
\eq
and the probability of microstates factorizes over sites, $P(\{m_i\})=f(m_1)\cdots f(m_N)$, with $f(m)=\exp(-c_2 m^{1/3})$. In this limit, our model corresponds to the B-in-B/ZRP model with a stretched-exponential weight function $f(m)$ \cite{evans0501338}. In particular, following \cite{b-in-b,evans0501338}, the critical density is given by
\bq
	\rho_c = \frac{F'(1)}{F(1)}, \label{rhoclf}
\eq
with 
\bq
	F(z) = \sum_{m=1}^\infty z^m \exp(-c_2 m^{1/3}) .
\eq
The above series does not admit a closed form, but it can be evaluated numerically for any $z$ and hence the critical density (\ref{rhoclf}) can be computed for any $c_2>0$. An important observable in this phase is the distribution of particles $p(m)$ - the probability that a randomly chosen node has $m$ particles. This corresponds to the distribution of three-volume in CDT. This distribution can be approximated as follows for $\rho\gg \rho_c$:
\bq
	p(m) \approx \exp(-c_2 m^{1/3})/F(1) + (1/N) p_{\rm cond}(m),
\eq
in which the first term corresponds to the critical distribution in the liquid bulk, and $p_{\rm cond}(m)$ denotes the probability of finding $m$ particles in the condensate. We can use the method of Ref.~\cite{b-in-b,majumdar0804-0197} to express this probability as follows:
\bq
	p_{\rm cond}(m) = N f(m) \frac{I(N-1,m,M-m)}{Z(N,M)}. \label{pcondlf}
\eq
Here $I(N-1,m,M-m)/Z(N,M)$ is the probability that the condensate has $m$ or less particles,
\bq
	I(N,m,M) = \sum_{m_1=1}^m \dots \sum_{m_N=1}^m \delta\left[M-\sum_i m_i\right] f(m_1) \dots f(m_N).
\eq
Following Ref.~\cite{majumdar0804-0197}, we replace the Delta function by its integral representation, and perform the sum over $\{m_i\}$. This gives 
\bq
	I(N-1,m,M-m) \approx \int_{-i\infty}^{i\infty} \frac{ds}{2\pi i} \exp\left[ N\left( \rho s - ms/N + \ln F(e^{-s})\right)\right].
\eq
The integral over $ds$ is dominated by its small-$s$ behaviour. We therefore expand $\ln F(e^{-s})$ at $s=0$,
\bq
	\ln F(e^{-s}) \cong \ln F(1) -s \frac{F'(1)}{F(1)} + \frac{s^2}{2} \left(\frac{F'(1)}{F(1)} - \frac{F'^2(1)}{F^2(1)} + \frac{F''(1)}{F(1)}\right) = \ln F(1) -s\rho_c + \frac{s^2}{2} \left(\rho_c - \rho_c^2 + \frac{F''(1)}{F(1)}\right),
\eq
and evaluate the resulting Gaussian integral. We obtain that the distribution of mass in the condensate (\ref{pcondlf}) is approximately Gaussian for $m$ close to $N(\rho-\rho_c)$:
\bq
		p_{\rm cond}(m) \propto \exp\left[-c_2 m^{1/3} - \frac{(m/N-(\rho-\rho_c))^2}{2(\rho_c - \rho_c^2 + \frac{F''(1)}{F(1)})}\right]. \label{pcondapprox}
\eq
This result agrees qualitatively with the simulations, 
see Fig.~\ref{fig:pcondlf}.

When $c_1<0$ and $c_2>0$, the condensate is still localised but the critical density is now zero, i.e., all particles go into the condensed phase. This so-called complete (or strong) condensation has its origin in the fact that the radius of convergence $z_c$ of $Z_N(z)$ from Eq.~(\ref{zcrit}) is becomes zero. This is because $T_{mn}(z)$ is unbound as either $m$ or $n$ approach infinity. The number of particles in the condensate is $\approx M$ and virtually does not fluctuate. The transition between the droplet phase and the localised phase is of second order, because the order parameters are continuous at $c_1=0$. 

\begin{figure}
\centering
\psfrag{x}{$k$} \psfrag{y}{$A(k)$}
\includegraphics[width=6cm]{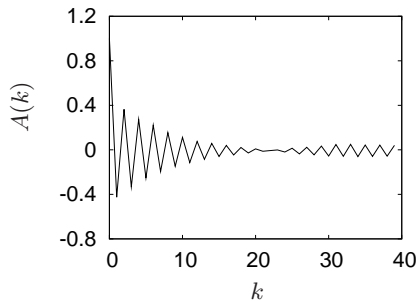}
\caption{\label{fig:corranti}Correlations in the antiferromagnetic phase obtained in MC simulations $N=80$, $M=18100$, $c_1=-0.5$, $c_2=-0.5$.}
\end{figure}

We shall now briefly discuss the antiferromagnetic phase. This phase exists in the region of both coupling constants being negative: $c_1<0,c_2<0$. The two-point weight $g(m,n)$ from Eq.~(\ref{gmn}) has now two positive terms: $(m-n)^2/(m+n)$ which prefers large differences in occupation numbers on neighbouring sites, and $m^{1/3}$ which prefers large occupations but itself does not lead to condensation. In Fig.~\ref{fig:corranti} we show the correlation function $A(k)$ for this phase. Its oscillatory behaviour reflects altered arrangement of occupied/empty sites. Interestingly, the correlation length is quite long, which may indicate a possible coupling between two neighbouring occupied sites via a not-completely-empty site between them.

Finally, let us consider the uncorrelated fluid phase which exists for $c_1=c_2=0$. The action $S[\{m_i\}]$ equals zero and the partition function can be calculated exactly:
\bq
	Z(N,M) = \sum_{m_1=1}^M ... \sum_{m_N=1}^M \delta\left(\sum_i m_i-M\right) = \binom{M-1}{M-N} \ .
\eq
We can now calculate the distribution of particles as follows (cf. Ref.~\cite{wbbj}):
\bq
	p(m) = \frac{Z(N-1,M-m)}{Z(N,M)} = \frac{(N-1)(M-N)!(M-m-1)!}{(M-1)!(M-m-N+1)!} \cong \frac{1}{\rho}\exp(-m/\rho),
\eq
where the last formula holds for $M\ll N$, i.e. for large density $\rho=M/N$ we typically deal with in this work. The distribution of particles (which corresponds to the distribution of three-volume) falls off exponentially with $m$.

\section{Uniqueness of $g(m,n)$}
\begin{figure}
	\psfrag{i}{$i$} \psfrag{mi}{$\bar m_i$} \psfrag{dmi}{$\sqrt{{\rm var} (m_i)}$}
	\includegraphics[width=6cm]{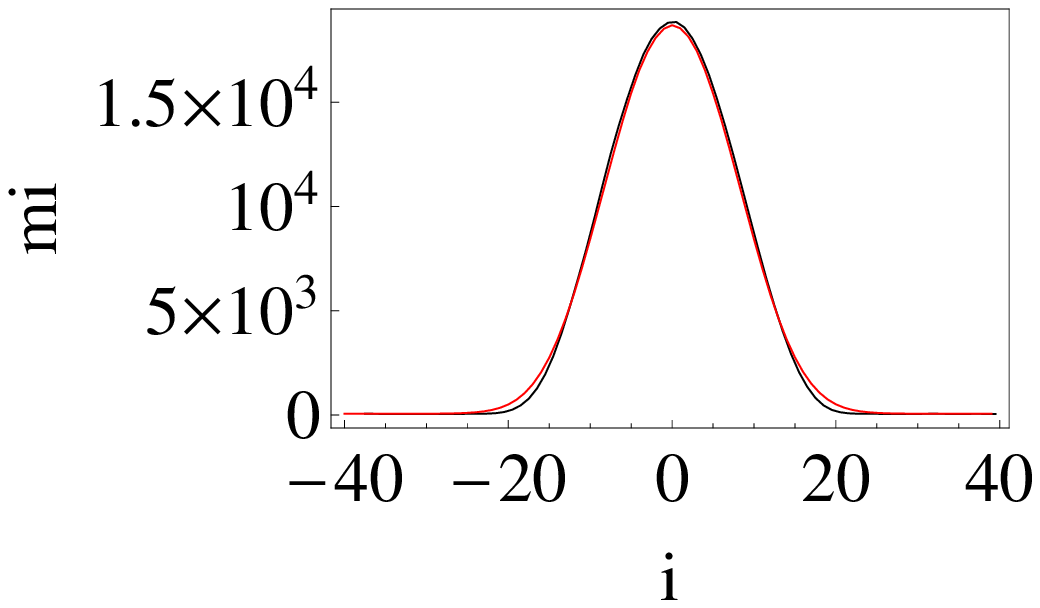} 
	\includegraphics[width=6cm]{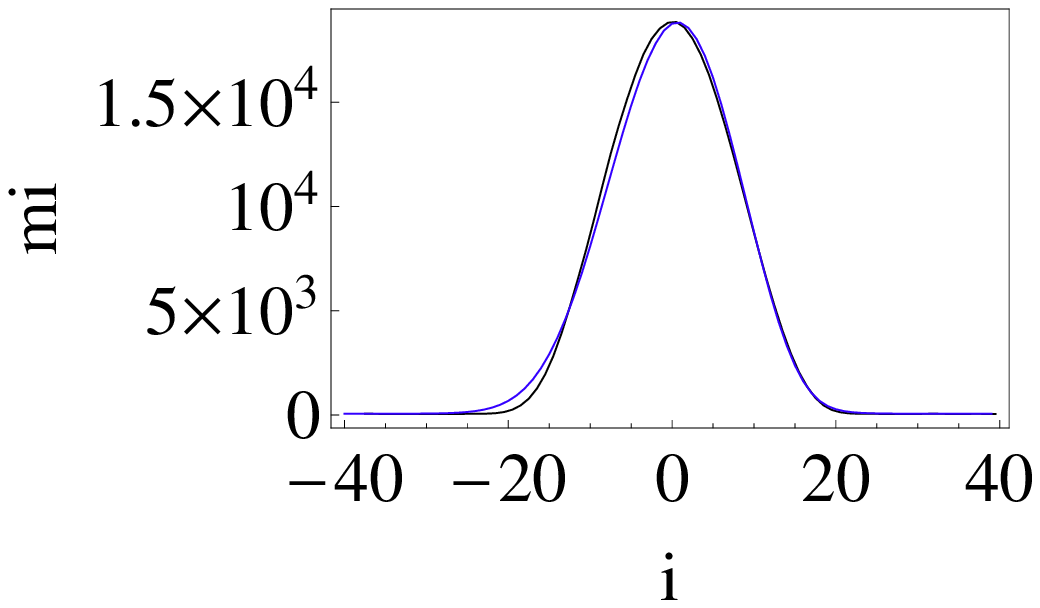} 
	\includegraphics[width=6cm]{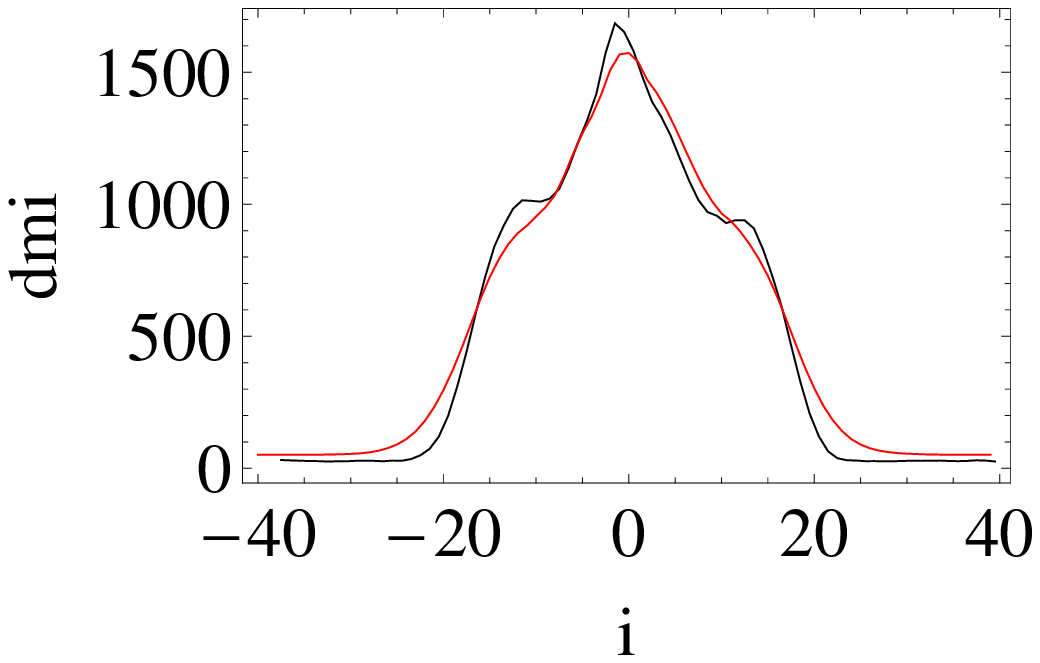}
	\includegraphics[width=6cm]{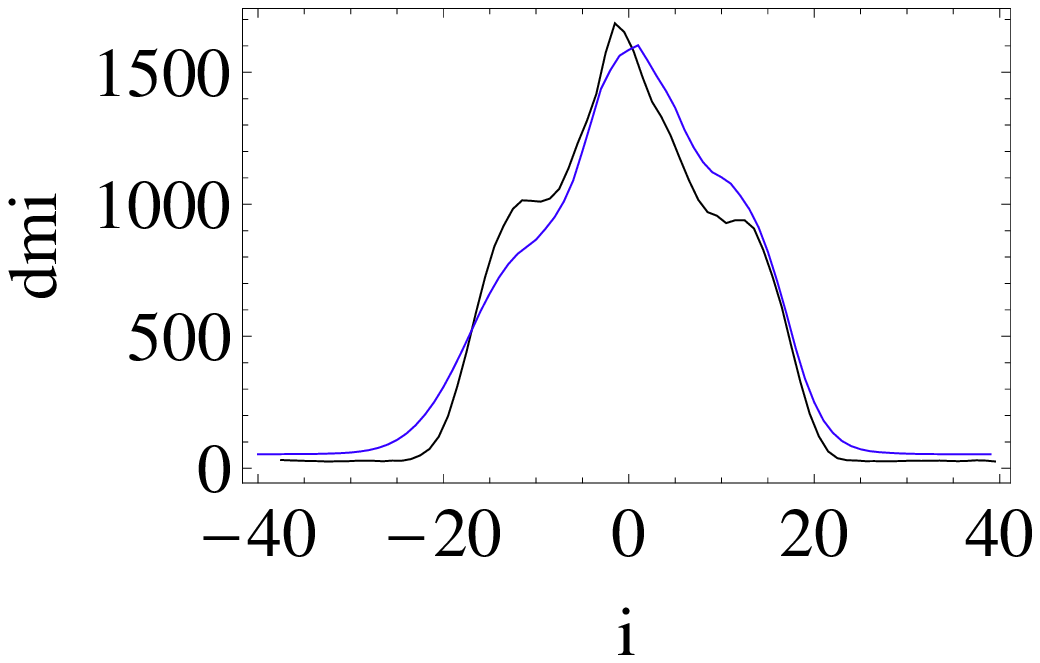}
	\caption{\label{sym_vs_asym}Comparison between the average droplet shape $\bar m_i$ (top) and quantum fluctuations $\sqrt{{\rm var} (m_i)}=\sqrt{\overline{m_i^2}-\overline{m}_i^2}$ (bottom) for asymmetric and symmetric $g(m,n)$, for $N=80, M=367200, c_1=0.01, c_2=0.59$. Black curves: MC simulations of the full CDT model (courtesy of A. G\"orlich) for $T=80$ time slices and the total volume $V_4=367200$ equivalent to the number of sites and particles in our simulations. Left: symmetric $g(m,n)$ from Eq.~(\ref{gmnsq}). The same result is obtained for the symmetric Eq.~(\ref{gmn}) studied in previous sections. Small asymmetry in $\sqrt{{\rm var} (m_i)}$ is caused by statistical fluctuations. Right: asymmetric $g(m,n)$ from Eq.~(\ref{gmnas}).}
\end{figure}

The choice of the transfer matrix $g(m,n)$ made in Eq. (\ref{gmn}) to reproduce the bell-shaped quantum universe is not unique. In fact, there is a whole family of functions $g(m,n)$ which lead to the following continuous limit:
\bq
	P(m_1,\dots,m_N) \to P(m(t)) = \exp\left[-\int dt \left(c_1\frac{m(t)'^2}{m(t)} +c_2 m(t)^{1/3}\right) \right],
\eq
and reproduce the shape given by Eq. (\ref{cos3}). In particular, two other forms of $g(m,n)$, the asymmetric one
\begin{equation}
g(m,n)=\exp{\left(-c_1 \frac{(m-n)^2}{m}-c_2 m^{1/3}\right)}, \label{gmnas}
\end{equation}
and the symmetric one with the geometric mean $\sqrt{mn}$ rather than the arithmetic mean $(m+n)/2$ in the denominator,
\begin{equation}
g(m,n)=\exp{\left(-c_1 \frac{(m-n)^2}{\sqrt{mn}} -c_2 \frac{m^{1/3} + n^{1/3}}{2}\right)} \ , \label{gmnsq}
\end{equation}
have the same asymptotic behaviour as Eq. (\ref{gmn}). Our simulations show (see Fig.~\ref{sym_vs_asym}) that the shape of the droplet is reproduced well by all three forms of $g(m,n)$ in the large-$N,M$ limit. However, the shape is slightly asymmetric in the case of Eq.~(\ref{gmnas}), whereas it is perfectly symmetric for symmetric forms of $g(m,n)$ as those given in Eqs.~(\ref{gmn}) or (\ref{gmnsq}). However, the data from the full CDT model are perfectly symmetric (excluding small statistical fluctuations). We thus conclude that the asymmetry is of finite-size origin and that the effective transfer matrix in CDT has to be symmetric as in Eqs.~(\ref{gmn}) or (\ref{gmnsq}). 
Interestingly, although Eq.~(\ref{gmnsq}) leads to exactly the same envelope (\ref{cos3}) in the droplet phase as Eq.~(\ref{gmn}), it does not permit the existence of the antiferromagnetic phase. Indeed, the corresponding action in the antiferromagnetic phase,
\bq
	S_{\rm antiferr.} \approx 2c_1 K^{-1/2} M^{3/2} + c_2 K^{2/3} M^{1/3},
\eq
is bigger than the corresponding action in the localised phase,
\bq
	S_{\rm localised} \approx 2c_1 M^{3/2} + c_2 M^{1/3},
\eq
for $c_1<0$ and for any $c_2$, and therefore antiferromagnetic states are disfavoured in this case. In other words,  the localised phase extends to all $c_2$ (positive and negative) in the phase plane (compare with Fig.~\ref{diagram_theor}) for the model with the transfer matrix given by Eq. (\ref{gmnsq}). We see that the existence 
of the antiferromagnetic phase depends on the behaviour of the kernel $g(m,n)$ for small values of the arguments.

\section{Conclusions}
In this paper we have analysed a simple model of particles residing on sites of a 1d lattice, in which the probability of microstate (\ref{Pss}) equals $e^{-S}$, where $S$ corresponds to the effective action (\ref{seff}) of the CDT model. We have shown that our model reproduces not only the average shape of the droplet -- the macroscopic universe of CDT -- but also quantum fluctuations around it. We have calculated the extension of this droplet and shown that the quantum universe is bigger than classical de-Sitter solution.

The droplet phase is one of five different phases which exists in our model. Two of these phases, localised condensate and uncorrelated fluid, can be identified as phases ``B'' and ``A'' of CDT. In each of these phases, we have calculated the distribution of particles $p(m)$ which corresponds to the distribution of three-volume in CDT. By measuring this distribution in the original CDT model and comparing it to our predictions one could validate  our hypothesis that all phases can be described by the same effective action.

Furthermore, we have predicted the existence of at least one more phase -- the correlated fluid phase. This phase, although yet unobserved, must surely exists in CDT as a simple consequence of periodic boundary conditions ensured by the global topology of CDT. We have calculated two observables: $p(m)$ and the correlation function $A(k)$, which can be easily measured in CDT. The agreement with our predictions would provide further evidence for the effective action (\ref{seff}).

Lastly, we have suggested that, depending on the behaviour of the action for small three-volumes, the fifth, antiferromagnetic phase can exist.

Our predictions can be tested in the CDT model, even without the knowledge of
the mapping between the effective coupling constants $c_1,c_2$ and the parameters in the Einstein-Hilbert action of CDT. In particular, the values of $c_1,c_2$ can be determined by fitting Eq.~(\ref{mt}) to the data from computer simulations in the macroscopic-universe phase, calculating $c_1/c_2$ from $W$, and resolving for $c_1,c_2$ using the equation for the background density $\rho_c$. Then, the correlated-fluid phase can be reached by increasing $M$. In other phases, equations derived in this paper for some quantities can be used to determine $c_1,c_2$. These values in turn can be applied to calculate other quantities and compare them to those estimated in full CDT simulations.

\section*{Acknowledgments}
We thank A. G\"orlich and J. Jurkiewicz for discussions and A. G\"orlich 
for providing us with data from CDT simulations. BW was supported by the EPSRC under grant EP/E030173 and ZB by the Polish Ministry of Science Grant No. N N202 229137 (2009-2012).

\section*{Appendix - Numerical Simulations}
Our model can be simulated using standard Monte Carlo techniques.
We start each simulation from some initial, random configuration of particles and construct a Markov chain in the space of configurations by moving particles between sites with probability depending on the current configuration. More specifically, we construct a new configuration $B=\{m_1,\dots,m_i-1,\dots,m_j+1,\dots,m_N\}$ from the old one $A=\{m_1,\dots,m_i,\dots,m_j,\dots,m_N\}$ by picking two random sites $i$ and $j$ with $m_i>1$, and moving one particle from site $i$ to site $j$ with probability given by the Metropolis formula
\bq
	P(A \rightarrow B)= \min{\left\{ 1,\frac{P(B)}{P(A)} \right\}} = \min{ \left\{1,\frac{g(m_{i-1},m_i-1)g(m_i-1,m_{i+1})g(m_{j-1},m_j+1)g(m_j+1,m_{j+1})}{g(m_{i-1},m_i)g(m_i,m_{i+1})g(m_{j-1},m_j)g(m_j,m_{j+1})} \right\}}
	\label{metropolis1}
\eq
if $i,j$ are not nearest neighbours, and with probability
\ba
	P(A \rightarrow B) = \min{ \left\{1,\frac{g(m_{i-1},m_i-1)g(m_i-1,m_{i+1}+1)g(m_{i+1}+1,m_{i+2})}{g(m_{i-1},m_i)g(m_i,m_{i+1})g(m_{i+1},m_{i+2})} \right\}} \quad \mbox{for} j=i+1, \label{metropolis2} \\
		P(A \rightarrow B) = \min{ \left\{1,\frac{g(m_{i-2},m_{i-1}+1)g(m_{i-1}+1,m_i-1)g(m_i-1,m_{i+1})}{g(m_{i-2},m_{i-1})g(m_{i-1},m_i)g(m_i,m_{i+1})} \right\}} \quad \mbox{for} j=i-1, \label{metropolis3} 
\ea
if they are neighbours, i.e., if $|i-j|=1$.
Such form of the acceptance probability guarantees that the probability of microstate $P(\{m_i\})$ will be given by Eq.~(\ref{Pss}). It is convenient to introduce the following notation:
\begin{equation}
\alpha(m,n)=\frac{g(m,n-1)}{g(m,n)}, \,\,\, \beta(m,n)=\frac{g(m-1,n)}{g(m,n)}, \,\,\, \gamma(m,n)=\frac{g(m-1,n+1)}{g(m,n)}, \,\,\, \delta(m,n)=\frac{g(m+1,n-1)}{g(m,n)}.
\end{equation}
Then, the acceptance probabilities can be rewritten as:
\ba
P(A \rightarrow B)=\min\left\{1, \frac{\alpha(m_{i-1},m_i) \beta(m_i, m_i+1)}{\alpha(m_{j-1},m_j+1) \beta(m_j+1,m_{j+1})} \right\}, \quad \mbox{for} \; |i-j|>1, \label{metropolis21} \\
P(A \rightarrow B)=\min\left\{1, \frac{\alpha(m_{i-1},m_i) \gamma(m_i, m_{i+1})}{\beta(m_{i+1}+1,m_{i+2})} \right\}, \quad \mbox{for} \; j=i+1, \label{metropolis22} \\
P(A \rightarrow B)=\min\left\{1, \frac{\beta(m_{i},m_{i+1}) \delta(m_{i-1}, m_{i})}{\alpha(m_{i-2},m_{i-1}+1)} \right\}, \quad \mbox{for} \; j=i-1. \label{metropolis23} \\
\ea
In our simulations, we calculate and store the values of $\alpha(m,n)$, $\beta(m,n)$, $\gamma(m,n)$, $\delta(m,n)$ for $m,n = 1,...,m_{\rm max}$, with some $m_{\rm max}<M$. This allows us to use Eqs.~(\ref{metropolis21})-(\ref{metropolis23}) and to avoid time-consuming computations of the ratios of $g(m,n)$ in Eqs.~(\ref{metropolis1})-(\ref{metropolis3}), if only the number of particles at sites $i,j$ does not exceed $m_{\rm max}$. Otherwise, we calculate the acceptance probability directly from Eqs.~(\ref{metropolis1})-(\ref{metropolis3}). The value of $m_{\rm max}$ - typically a few thousands - is chosen as big as possible given available computer memory. To reduce the autocorrelation time, measurements are made every $M$ moves. 

All measurements of the average shape of the condensate and fluctuations around it are performed by shifting the condensate for each sample to a common centre of mass at site $i=N/2$. In order to account for periodic boundary conditions, the centre of mass is found in a 2d plane, assuming that the sites reside on a circle in this plane, and then the coordinates $(x,y)$ of that point are mapped to the index $i$ of a site closest to the centre of mass. We have checked that other procedures of finding the centre lead to very similar results.

\end{document}